\documentclass[journal]{IEEEtran}
\usepackage{amsfonts}
\usepackage{amsmath}
\usepackage{cite}
\usepackage{array}
\usepackage{mdwmath}
\usepackage{mdwtab}
\usepackage{graphicx,epsfig}
\usepackage{eqparbox}
\usepackage{subfigure}
\usepackage{caption}
\usepackage{url}
\usepackage{amssymb}
\usepackage{float}
\usepackage{indentfirst}
\usepackage{tabularx,ragged2e}
\usepackage{cases}
\usepackage{algorithm}
\usepackage{algorithmic}
\usepackage{mathtools}
\usepackage{bm}
\usepackage{setspace}
\usepackage{makecell}
\usepackage{xcolor}
\usepackage{ntheorem}
\usepackage{epstopdf}
\usepackage{verbatim}
\usepackage{enumerate}
\usepackage{color}
\usepackage{subfigure}
\theoremheaderfont{\bfseries}
\theorembodyfont{\upshape}
\theoremseparator{:}
\newtheorem{theorem}{Theorem}
\newtheorem{lemma}{Lemma}

\newtheorem{corollary}{Corollary}


\newcolumntype{C}{>{\Centering\arraybackslash}X}

\begin{document}
\title{Unsourced Random Massive Access with Beam-Space Tree Decoding}
\author{
Jingze~Che,~\IEEEmembership{Student~Member,~IEEE, }
Zhaoyang~Zhang,~\IEEEmembership{Senior~Member,~IEEE, }
Zhaohui~Yang,~\IEEEmembership{Member,~IEEE, }
Xiaoming~Chen,~\IEEEmembership{Senior~Member,~IEEE, }
Caijun~Zhong,~\IEEEmembership{Senior~Member,~IEEE, }
and~Derrick~Wing~Kwan~Ng,~\IEEEmembership{Fellow,~IEEE} 

\thanks{The work of J. Che, Z. Zhang, X. Chen and C. Zhong was supported in part by National Key R\&D Program of China under Grant 2018YFB1801104 and 2020YFB1807101, and National Natural Science Foundation of China under Grant 61725104, 61922071 and U20A20158. The work of D. W. K. Ng was supported in part by funding from the UNSW Digital Grid Futures Institute, UNSW, Sydney, under a cross-disciplinary fund scheme and by the Australian Research Council's Discovery Project (DP210102169).}
\thanks{J.~Che, Z.~Zhang (Corresponding Author), X. Chen and C. Zhong (e-mails: \{jzche, ning\_ming, chen\_xiaoming, caijunzhong\}@zju.edu.cn) are with 1) College of Information Science and Electronic Engineering, Zhejiang University, Hangzhou 310027, China, and 2) International Joint Innovation Center, Zhejiang University, Haining 314400, China, and also 3) Zhejiang Provincial Key Laboratory of Info. Proc., Commun. \& Netw. (IPCAN), Hangzhou 310027, China. }
\thanks{Z.~Yang (e-mail: zhaohui.yang@ucl.ac.uk) is with the Department of Electronic and Electrical Engineering, University College London, UK, and he is also a visiting scholar at Zhejiang University.}
\thanks{D.~W.~K.~Ng (e-mail: w.k.ng@unsw.edu.au) is with the
School of Electrical Engineering and Telecommunications, UNSW, Sydney, Australia.}
}

\maketitle

\begin{abstract}
    The core requirement of massive Machine-Type Communication (mMTC) is to support reliable and fast access for an enormous number of machine-type devices (MTDs). In many practical applications, the base station (BS) only concerns the list of received messages instead of the source information, introducing the emerging concept of unsourced random access (URA). Although some massive multiple-input multiple-output (MIMO) URA schemes have been proposed recently, the unique propagation properties of millimeter-wave (mmWave) massive MIMO systems are not fully exploited in conventional URA schemes. In grant-free random access, the BS cannot perform receive beamforming independently as the identities of active users are unknown to the BS. Therefore, only the intrinsic beam division property can be exploited to improve the decoding performance. In this paper, a URA scheme based on beam-space tree decoding is proposed for mmWave massive MIMO system. Specifically, two beam-space tree decoders are designed based on hard decision and soft decision, respectively, to utilize the beam division property. They both leverage the beam division property to assist in discriminating the sub-blocks transmitted from different users. Besides, the first decoder can reduce the searching space, enjoying a low complexity. The second decoder exploits the advantage of list decoding to recover the miss-detected packets. Simulation results verify the superiority of the proposed URA schemes compared to the conventional URA schemes in terms of error probability.
\end{abstract}

\IEEEpeerreviewmaketitle
\section{Introduction}
\subsection{Motivation}
\IEEEPARstart{O}{ne} imminent demand for the next generation wireless mobile communication systems is to provide instant and reliable access for an increasingly large number of machine-type devices (MTDs) \cite{bockelmann2016massive, shariatmadari2015machine}. Different from human-centric communication, the resultant Massive Machine-Type Communication (mMTC) has two distinct features. In particular, only a small number of devices are active in each communication round due to the sporadic activity in mMTC\cite{chen2018sparse}. Besides, MTDs usually transmit small data payloads adopting short-packet signaling \cite{durisi2016toward}. These make traditional grant-based random access schemes generally not very suitable for the mMTC scenario because of their low spectral efficiency and exceedingly long latency \cite{chen2020massive}. Therefore, the design of reliable and efficient grant-free random access schemes has attracted significant attention recently, where active users transmit pilots and data to the base station (BS) directly without permission granted \cite{liang2017non,zhang2016grant}. In most grant-free random access schemes, a set of pilot sequences that are designated to the users are used for the BS to ensure its accurate user activity detection and channel estimation \cite{senel2018grant, kim2019novel, wang2019joint}. However, this is neither affordable nor feasible in the next generation multiple access (NGMA) scenarios due to the high density, the large number of connections therein, and the frequent collisions that may occur. To tackle the issues, a special type of grant-free random access, the so-called unsourced random access (URA), is introduced in \cite{polyanskiy2017perspective}, in which users do not transmit preambles, all the potential users share a common codebook, and the BS only needs to decode a list of messages instead of the identities of active users. This scheme can avoid the huge cost of preambles and the extra protocol of collision resolution, thus well meeting the requirements of next generation massive access. 

On the other hand, massive or super MIMO technology, in combination with the millimeter-wave (mmWave) technology, have been promoted as two core technological features for the next generation wireless communication system with a witnessed potential to boost the capacity and efficiency. These two underlying key technologies jointly bring additional spatial-domain signal dimension with their excellent intrinsic directivity and proper beamforming, and also result in salient beam-space sparsity due to the lack of scattering in a mmWave MIMO channel \cite{akdeniz2014millimeter,wen2014channel,bellili2019generalized}. 
To further increase the efficiency of the future massive access systems, such spatial-domain resources and properties should be fully explored and exploited. Various multi-user transmission schemes have been proposed to unleash the potential and properties of the beam-space resources, such as the typical works on beam division multiple access (BDMA), which simultaneously serves multiple users via different beams \cite{sun2015beam, you2017bdma, jia2019massive}. 
    
However, for grant-free random access, even if the information of the location of all potential users is stored at the BS, the identities of active users are unknown to the BS. Therefore, the beam dimension cannot be directly exploited in the design of the encoding process of an unsourced grant-free random access, which is different from previous works \cite{sun2015beam, you2017bdma, jia2019massive}. Moreover, in a general URA system, the messages of active users are divided into several sub-blocks and transmitted in consecutive sub-slots. As the signals transmitted from active users often experience deep fading, some sub-blocks may be missed by the decoder at the receiver. The loss of any sub-block in any sub-slot finally leads to the failure of recovering the corresponding original message. The problem of packet loss is also needed to be solved as it can cause severe decoding performance degradation. Note that the user location information can generally serve as a hint of the indices of the messages originated from it. Therefore, the intrinsic beam division property and salient spatial sparsity in a mmWave MIMO system can provide extra extrinsic information for both multi-user signal separation and multi-sub-block message stitching. This motivates us to exploit the beam space properties to design new URA schemes for next generation multiple access to help the entire system accommodate more active users and improve the decoding performance. 

\subsection{Related Works}
Y. Polyanskiy first introduced a framework named URA \cite{polyanskiy2017perspective}. Specifically, in URA, all the users share a common codebook, and the decoder only needs to decode a list of messages transmitted from the active users. The error probability is defined as the average fraction of mis-decoded messages over the number of active users, including both missed detection and false alarm. It is obvious that the message recovery at the BS can be formulated as a compressed sensing (CS) problem due to the sporadic activity in mMTC, which is similar to the conventional grant-free random access schemes \cite{liu2018massive, liu2018massive2}. However, the size of the common codebook grows exponentially with the number of information bits. In practice, even if a short packet is transmitted, the size of information messages is typically at the order of 100 bits, which makes the CS algorithms computationally intractable. In this context, V. K. Amalladinne $et \; al.$ proposed a coded compressed sensing (CCS) scheme for URA communication \cite{amalladinne2020coded}. In particular, the messages from active users are first divided into several sub-blocks. Then, a systematic linear code adds redundancy to those sub-blocks. Once this is achieved, each sub-block is mapped into a codeword in a common codebook and transmitted in a certain sub-slot.  Then a standard CS algorithm implements the detection of the sub-blocks. Finally, the sub-blocks transmitted in different sub-slots are stitched together to obtain the original messages. Build upon the findings in \cite{amalladinne2020coded} and the structure of sparse regression codes (SPARCs), A. Fengler $et\;al.$ provided an improved inner decoder, and a complete asymptotic error analysis \cite{fengler2021sparcs}.

Apart from the above works, the study of massive multiple-input multiple-output (MIMO) URA has also attracted much attention. A. Fengler $et\;al.$ extended the URA model of \cite{amalladinne2020coded} to a block-fading MIMO channel by using a low-complexity covariance-based CS (CB-CS) recovery algorithm \cite{fengler2021non}. Considering the low code rate and spectral efficiency of the CCS scheme, V. Shyianov $et\;al.$ proposed a new algorithmic solution to solve the massive URA problem by leveraging the rich spatial dimensionality offered by large-scale antenna arrays \cite{shyianov2020massive}. Besides, without requiring a separate activity detection or channel estimation step, A. Decurninge $et\;al.$ introduced a structure that allows the receiver to separate the users using a classical tensor decomposition \cite{decurninge2020tensor}. As URA is a special scheme of grant-free random access, A. Fengler $et\;al.$ presented a conceptually simple algorithm based on pilot transmission, activity detection, channel estimation, Maximum Ratio Combining (MRC), and single-user decoding \cite{fengler2020pilot}, which is similar to the existing grant-free random access schemes \cite{chen2018sparse, liu2018massive}. The difference is that they use a pool of non-orthogonal pilots where every active user picks one of them pseudo-randomly. Furthermore, X. Shao $et\;al.$ proposed a unified cooperative activity detection framework for sourced and unsourced random access based on the covariance of the received signals for the sixth generation (6G) cell-free wireless networks \cite{CAD}.

\subsection{Main Contributions}
In this paper, we propose a URA scheme with beam-space tree decoding. Specifically, we adopt the CCS scheme \cite{amalladinne2020coded} suitably to our case and design two beam-space tree decoders, which are based on hard decision and soft decision, respectively.
By leveraging the beam division property to assist in distinguishing the sub-blocks transmitted from different users, both decoders can help the system serve more active users. As the discriminating power is improved, the searching space of the solution in the decoding process is reduced, such that the first decoder has low complexity. In addition, notice that any sub-block missed by the CS decoder would finally lead to missed detection, which degrades the decoding performance. To tackle this issue, the second decoder establishes factor graphs at each stage during the decoding process and implements message passing algorithm (MPA) to give each candidate sub-block drawn from the checking relationship a log-likelihood ratio (LLR) value. Then the reliability of every candidate path is calculated by a path metric (PM). At every stage, some reliable paths are kept, and finally, the surviving path is output as the valid message. Even if a sub-block is missed by the CS decoder, it is possible that the path of the original message is reliable and kept.
The main contributions of this paper are summarized as follows:
\begin{enumerate}
\item[$\bullet$] A URA scheme with beam-space tree decoding is proposed for mmWave communication systems in mMTC to accommodate more active users and to improve the system performance.
\item[$\bullet$] Two beam-space tree decoders are designed. Both of them can exploit the intrinsic beam division property to improve the decoding performance of the tree decoder by enhancing the discriminating power and helping the system serve more active users. Besides, the first decoder is based on hard decision with low complexity. The second one is based on soft decision and exploits the advantage of list decoding to recover the packet loss, which is the key of the proposed URA scheme.
\item[$\bullet$] Simulation results verify that our URA schemes have significantly better performances than existing works.
\end{enumerate}

\subsection{Paper Organization and Notations}
The rest of this paper is organized as follows: Section II provides a brief introduction of the considered  massive URA system. Section III provides the encoding and decoding process of the considered system. Section IV proposes a beam-space tree decoder with hard decision. Then, Section V designs a beam-space tree decoder with soft decision. Next, Section VI analyzes the performance of the proposed URA scheme. Afterward, Section VII provides extensive simulation results to validate the effectiveness of the proposed algorithm. Finally, Section VIII concludes the paper.

Throughout this paper, we use bold letters to denote matrices or vectors and non-bold letters to denote scalars. We denote the $i$-th row and the $j$-th column of a matrix $\mathbf{X}$ with the row-vector $\mathbf{X}_{i,:}$ and the column-vector $\mathbf{X}_{:,j}$ respectively. We denote $\mathbb{C}^{A\times B}$ by the space of complex matrices of size $A\times B$. We use $|\cdot|$ to denote the absolute value of a complex number, $(\cdot)^{\rm{H}}$ and $(\cdot)^{\rm{T}}$ to denote conjugate transpose and transpose, respectively. The $l_i$-norm of an input vector is denoted by $\left\| {\cdot} \right\|_i$. $\left|{\mathcal{K}}\right|_{\mathrm{c}}$ denotes the number of elements of set $\mathcal{K}$. The notation $x \sim \mathcal{CN}(\mu, \delta^{2})$ denotes that the random variable (r.v.) $x$ follows the circular symmetric complex Gaussian distribution. $\mathcal{O}(\cdot)$ stands for the big-O notation.
\begin{figure*}[!htp]
\centering
\includegraphics[width=0.9\linewidth]{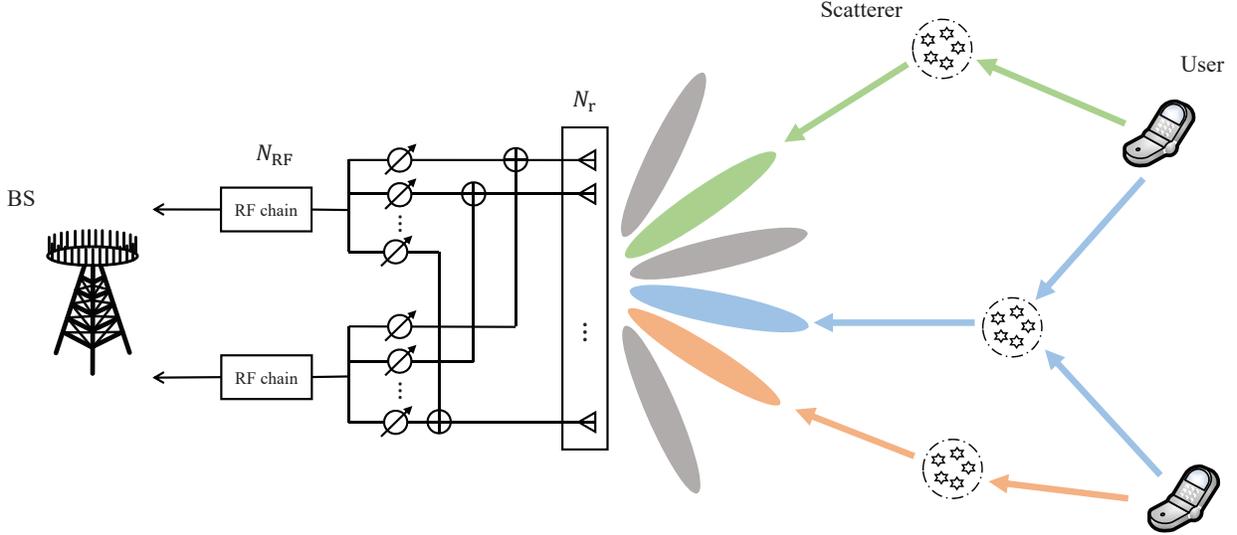}
\caption{$\mathrm{The\;system\;model\;of\;our\;proposed\;scheme.}$}
\label{system_model}
\end{figure*}
\section{System Model}
Consider an uplink single-cell cellular network consisting of $K_{\mathrm{total}}$ single-antenna users. The BS is equipped with $N_{\mathrm{r}}$ antennas and $N_{\mathrm{RF}}$ radio frequency (RF) chains such that $N_{\mathrm{RF}} < N_{\mathrm{r}}$, as shown in Fig. \ref{system_model}. Due to the sporadic user activity of mMTC, only a small number of $K_{\mathrm{a}}$ users are active in a transmisson process, i.e., $K_{\mathrm{a}} \ll K_{\mathrm{total}}$. Each active user has $B$ bits of information to be transmitted in a block-fading channel. According to \cite{akdeniz2014millimeter, bellili2019generalized}, the channel vector $\mathbf{h_k}$ from user $k$ to the BS can be written as
\begin{equation}
{\mathbf{h}_k} = \sum\limits_{p=1}^{P_c} \sum\limits_{q = 1}^{Q_p} {\beta_{p,q}\mathbf{e}(\theta_{p,q})},
\end{equation}
where $P_c$ denotes the total number of clusters and within the $p$-th cluster there are $Q_p$ sub-paths.
$\beta_{p,q}$ and $\theta_{p,q}$ denote the gain and the the angle of arrival (AOA) of the $q$-th sub-path within the $p$-th cluster. For the uniform linear array (ULA), the $N_{\mathrm{r}} \times 1$ array steering vector $\mathbf{e}(\theta)$ can be expressed as
\begin{equation}
	\mathbf{e}(\theta ) = {[{e^{ - \frac{{j2\pi d\sin (\theta )m}}{\lambda }}}]}_{m \in J(N_{\mathrm{r}})},
\end{equation}
where $J(N_{\mathrm{r}})=\{i-\frac{N_{\mathrm{r}}-1}{2},i=0,1,2,\ldots,N_{\mathrm{r}}-1\}$, $\theta \in [-\frac{\pi}{2}, \frac{\pi}{2}]$, $\lambda$ is the signal wavelength, and $d$ is the antenna spacing which is usually half of the signal wavelength.

To overcome the strong path loss in mmWave channels, a beamforming technique should be adopted. 
However, the BS cannot focus in any specific direction in grant-free random access. The reason is that even if the location of all potential users is stored at the BS, which users are active is not prior information known to the BS.
Besides, due to the constraint of hardware implementations and large energy consumption of RF chains, we have $N_{\mathrm{RF}}<N_{\mathrm{r}}$. Therefore, many beamforming methods in existing works \cite{gao2016near,wang2017spectrum} cannot be applied in our system directly as $N_{\mathrm{RF}}$ narrow beams cannot cover the whole beam space. In this paper, we give a beamforming method based on the widely used Discrete Fourier Transform (DFT) based beamforming codebook \cite{gao2016near,wang2017spectrum}, to overcome the strong path loss of mmWave channels in grant-free random access.
Specifically, the DFT based beamforming codework, which is denoted by $\mathbf{W}$, can be writtern as
\begin{equation}
  \mathbf{W} = [{\mathbf{w}_1},{\mathbf{w}_2},\ldots,{\mathbf{w}_{N_{\mathrm{r}}}}] \in {\mathbb{C}^{{N_{\mathrm{r}}} \times {N_{\mathrm{r}}}}},
  \end{equation}
where
\begin{equation}
\begin{aligned}
  \mathbf{w}_i &= \frac{1}{\sqrt{N_{\mathrm{r}}}}\mathbf{e}(\theta_i), \\
  {\theta _i} &= \arcsin (\frac{{2i - 1}}{{{N_{\mathrm{r}}}}} - 1),\;i = 1,2,\ldots,{N_{\mathrm{r}}}. 
\end{aligned}
\end{equation}
Consider the process of hardware implementations, the number of antennas is usually a multiple of the number of RF chains. Therefore, the $\frac{N_{\mathrm{r}}}{N_{\mathrm{RF}}}$ consecutive beamforming vectors can be grouped and summed together to form a new beamforming vector $\mathbf{\overline{w}}_i$, $i=1,2,\ldots,N_{\mathrm{RF}}$. $\mathbf{\overline{w}}_i$ is expressed as
\begin{equation}
  \mathbf{\overline{w}}_i=\gamma(\mathbf{w}_{1+\frac{N_{\mathrm{r}}}{N_{\mathrm{RF}}}(i-1)}+\mathbf{w}_{2+\frac{N_{\mathrm{r}}}{N_{\mathrm{RF}}}(i-1)}+ \cdots +\mathbf{w}_{\frac{N_{\mathrm{r}}}{N_{\mathrm{RF}}}i}),
\end{equation}
where the parameter $\gamma$ is set to constrain the power of receive beamforming, i.e., $\| \mathbf{\overline{w}}_i \|^2_2=1$. Then the beamforming matrix $\mathbf{\overline{W}}$ is obtained, where $\mathbf{\overline{W}}$ is written as $\mathbf{\overline{W}}=[\mathbf{\overline{w}}_1,\mathbf{\overline{w}}_2,\ldots,\mathbf{\overline{w}}_{N_{\mathrm{RF}}}] \in \mathbb{C}^{N_{\mathrm{r}} \times N_{\mathrm{RF}}}$. By applying this beamforming method, the width of every beam is $\frac{\pi}{N_{\mathrm{RF}}}$, thus the $N_{\mathrm{RF}}$ beams can cover the whole beam space, which means that the signals coming from all directions can be received by the BS.

In a typical URA scenario, all the users share a common codebook $\mathbf{A}$, which is denoted by $\mathbf{A}=[\mathbf{a}_1, \mathbf{a}_2,\ldots,\mathbf{a}_{N}] \in {\mathbb{C}^{{L_{\mathrm{p}}} \times {N}}}$. The power of each codeword $\mathbf{a}_n \in {\mathbb{C}^{L_{\mathrm{p}} \times 1}}$ is constrained to 1, i.e., $\left\| {{\mathbf{a}_n}} \right\|_2^2 = 1$. Let $\delta_{n,k} \in \{0,1\}$ denote whether user $k$ transmits the codeword $\mathbf{a}_n$. $\delta_{n,k}$ can be written as
\begin{equation}
{\delta _{n,k}} = \left\{ {
\begin{array}{*{20}{c}}
1,&\mathrm{active\;user}\;k\;\mathrm{transmits\;codeword}\;{\mathbf{a}_n},\\
0,&\mathrm{otherwise}.
\end{array}
} \right.
\end{equation}

After receive beamforming at the receiver, the beam domain channel vector of the active user $k$ is denoted by ${\overline{\mathbf{h}}_k} = {\mathbf{\overline{W}}^{\rm{H}}}{\mathbf{h}_k}=[\overline{h}_{k,1},\overline{h}_{k,2},\ldots,\overline{h}_{k,N_{\mathrm{RF}}}]^{\rm{T}}$. Also, the random noise vector is denoted by $ {\overline{\mathbf{z}}} = {\mathbf{\overline{W}}^{\rm{H}}}{\mathbf{z}}=[\overline{z}_{1},\overline{z}_{2},\ldots,\overline{z}_{N_{\mathrm{RF}}}]^{\rm{T}}$, where $\mathbf{z}$ is modeled by a complex circular Gaussian random vector with i.i.d. components, i.e., ${\mathbf{z}} \sim \mathcal{CN}(0,\sigma_z^2\mathbf{I})$. Then the received signal on the $b$-th beam can be written as
\begin{equation}
{\mathbf{y}_b} = \sum\limits_{k = 1}^{{K_{\mathrm{total}}}} {\sum\limits_{n = 1}^N {{{\overline{h}}_{k,b}}{\delta _{n, k}}\mathbf{a}_n^{\rm{T}} + {{\overline{\mathbf{z}}}_b}} },
\end{equation}

By summarizing all the $N_{\mathrm{RF}}$ samples in a transmission block, the received signal can be recast as
\begin{equation}
\mathbf{Y} = \mathbf{A}\mathbf{\Delta}\overline{\mathbf{H}} + \overline{\mathbf{Z}} = \mathbf{A}\overline{\mathbf{X}} + \overline{\mathbf{Z}},
\end{equation}
\begin{figure}
\begin{center}
\subfigure[]{
\includegraphics[width=0.57\linewidth,height=2.7cm]{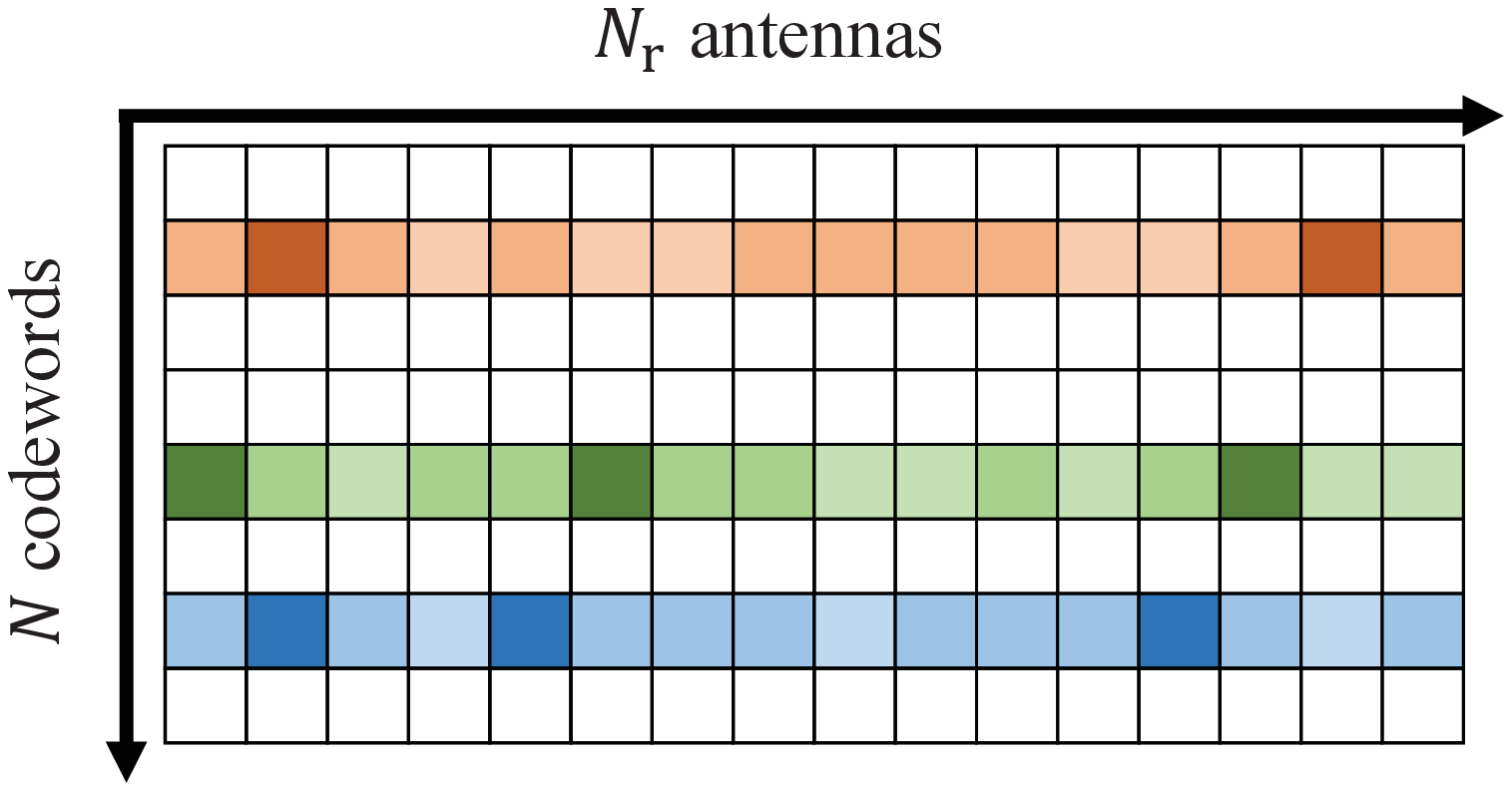}
\label{antenna_code}
}
\subfigure[]{
\includegraphics[width=0.33\linewidth,height=2.7cm]{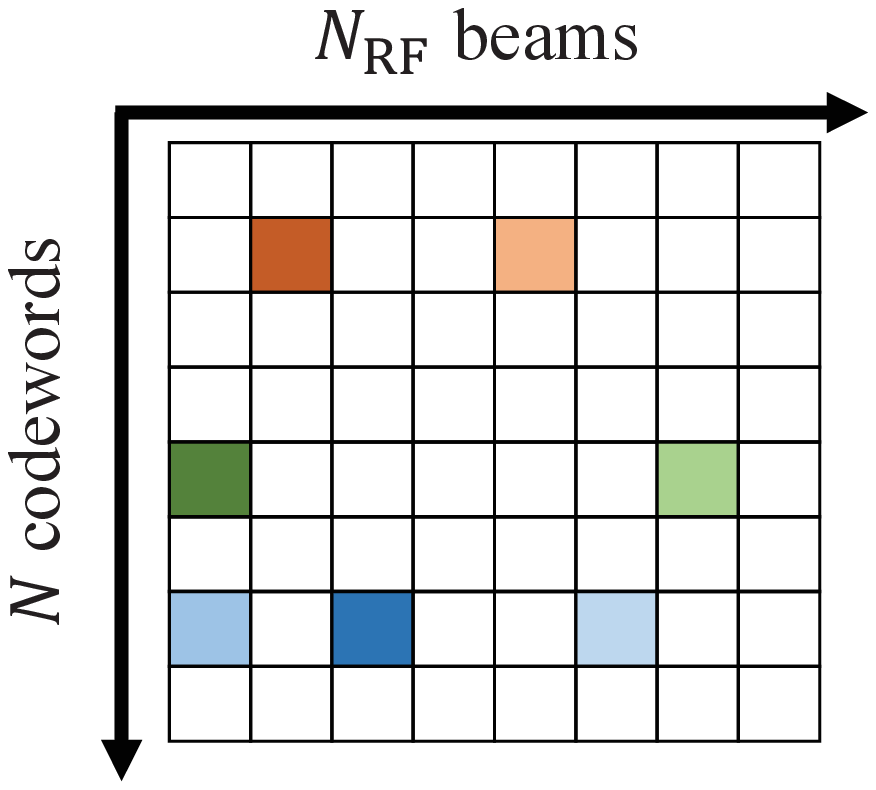}
\label{beam_code}
}
\caption{\rm (a) The matrix $\mathbf{X}$ exhibits the sparsity of user activity; (b) The matrix $\overline{\mathbf{X}}$ exhibits the sparsity of beam domain channel in mmWave bands.}
\end{center}
\end{figure}
where $\mathbf{Y} = [\mathbf{y}_1^{\rm{T}},\mathbf{y}_2^{\rm{T}},\ldots,\mathbf{y}_{{N_{\mathrm{RF}}}}^{\rm{T}}] \in {\mathbb{C}^{{L_{\mathrm{p}}} \times {N_{\mathrm{RF}}}}}$, $\overline{\mathbf{H}} = [\overline{\mathbf{h}}_1,\overline{\mathbf{h}}_2,\ldots,$ $\overline{\mathbf{h}}_{{K_{\mathrm{total}}}}]^{\rm{T}} \in {\mathbb{C}^{{K_{\mathrm{total}}} \times {N_{\mathrm{RF}}}}}$,
$\overline{\mathbf{X}} = \mathbf{\Delta}\overline{\mathbf{H}} \in {\mathbb{C}^{N \times {N_{\mathrm{RF}}}}}$ and $\mathbf{\Delta}=\{0,1\}^{N \times K_{\mathrm{total}}}$. The matrix $\mathbf{\Delta}$ contains only $K_{\mathrm{a}}$ non-zero columns each of which having a non-zero entry. 

For the matrix ${\mathbf{X}}=\mathbf{\Delta}{\mathbf{H}}$, where $\mathbf{H}=[\mathbf{h}_1,\mathbf{h}_2,\ldots,$ $\mathbf{h}_{K_{\rm{total}}}]^{\rm{T}}$, the $i$-th row of such matrix is given as
\begin{equation}
    {\mathbf{X}}_{i,:}=\sum\limits_{k=1}^{K_{\rm{total}}}\delta_{i,k}{\mathbf{h}}_k^{\rm{T}}.
\end{equation}
The probability that ${\mathbf{X}}_{i,:}$ is identically zero is given by $(1-2^{-J})^{K_{\mathrm{a}}}$. Since $2^J$ is significantly larger than $K_{\mathrm{a}}$, the matrix ${\mathbf{X}}$ is row-sparse, which is shown in Fig. \ref{antenna_code}. The reason is that only a small number of users are active due to the sporadic traffic of users, i.e., $K_{\mathrm{a}} \ll K_{\rm{total}}$. For the same reason, the matrix $\overline{\mathbf{X}}$ is also row-sparse. Moreover, due to the lack of scattering in mmWave bands, the signal propagates from the transmitter to the receiver through a small number of path clusters. This leads to the sparsity of mmWave massive MIMO channels in the beam domain as well, i.e., the channel vector $\overline{\mathbf{h}}_k$ is sparse. Therefore, for the matrix $\overline{\mathbf{X}}$, the number of non-zero entries of its columns is less than that of ${\mathbf{X}}$, which is shown in Fig. \ref{beam_code}. Note that the total number of users $K_{\rm{total}}$ plays no role in the matrix $\overline{\mathbf{X}}$. This means that if the matrix $\overline{\mathbf{X}}$ is recovered, only the codeword index is known to the BS, instead of the user's ID, which leads to the so-called unsourced property.

Let $\mathcal{L}$ and $\mathcal{K}_{\mathrm{a}}$ denote the set of the recovered messages at the BS and the set of the active users, respectively. Each active user $k \in \mathcal{K}_{\mathrm{a}}$ expects to transmit $B$ bits of information, i.e., ${\mathbf{u}_k} = {\{ 0,1\} ^B}$. The performance in URA is evaluated by the probability of missed detection and false alarm, denoted by $p_{\mathrm{md}}$ and $p_{\mathrm{fa}}$ respectively, which can be given by:
\begin{equation}
{p_{\mathrm{md}}} = \frac{1}{{{K_{\mathrm{a}}}}}\sum\limits_{k \in \mathcal{K}_{\mathrm{a}}} {\mathrm{Pr}({\mathbf{u}_k} \notin \mathcal{L})},
\end{equation}
\begin{equation}
{p_{\mathrm{fa}}} = \frac{{\left| {\mathcal{L}\backslash \{ {\mathbf{u}_k}|k \in \mathcal{K}_{\mathrm{a}}\} } \right|_{\mathrm{c}}}}{{\left| \mathcal{L} \right|_{\mathrm{c}}}},
\end{equation}
and the error probability of the system is defined as
\begin{equation}
p_{\mathrm{err}}=p_{\mathrm{md}}+p_{\mathrm{fa}}.
\end{equation}

\section{Proposed URA Scheme}
In this section, we first review the studies of the CCS scheme in \cite{amalladinne2020coded} and then propose a URA scheme. In the CCS scheme, each active user partitions the message into several sub-blocks and adds parity bits. The CS techniques detect the sub-blocks transmitted by active users in all sub-slots. A tree-based algorithm then stitched these sub-blocks to recover the original messages. 
\subsection{Encoding Process}
The transmission strategy includes two encoders: tree encoder and CS encoder. The tree encoder uses a systematic linear block code based on random parity checks to add parity bits to every sub-block. The CS encoder maps each sub-block into a codeword in the common codebook.
\subsubsection{Tree Encoder}
 Divide $B$ bits message into $S$ sub-blocks of size $b_1,b_2,\ldots,b_S$, where $\sum\nolimits_{i = 1}^S {{b_i}}  = B$. Let $b_1=J$ and $b_s < J$, $s = 2,3,\ldots,S$. Each sub-block $s$ is resized to length $J$ by appending $l_s=J-b_s$ parity bits, which is obtained by linear combinations of the information bits of the previous sub-blocks. Mathematically, define $\mathbf{m}$ as a coded message, then we have $\mathbf{m} =\left[\mathbf{m}(1),\mathbf{m}(2),\dots,\mathbf{m}(S)\right]= \left[\mathbf{b}(1) , \mathbf{l}(1) , \mathbf{b}(2) , \mathbf{l}(2) , \dots ,  \mathbf{b}(S) , \mathbf{l}(S) \right]\in\{0,1\}^{1 \times JS}$. Herein, ${\mathbf{l}}(s)$ is obtained by
\begin{equation}
{\mathbf{l}}(s) = \sum\limits_{i = 1}^{s-1} {{\mathbf{b}}(i){\mathbf{G}_{i,s-1}}},
\label{parity}
\end{equation}
where $\mathbf{G}_{i,s-1} \in \{0,1\}^{b_i \times l_s}$ is a binary matrix. Parity bits are computed using modulo-2 arithmetic and, as such, they remain binary. Every sub-block has the same size $b_s + l_s = J$, and the code rate $R_{\mathrm{tree}}$ is fixed as $R_{\mathrm{tree}} = \frac{B}{{JS}}$.
\subsubsection{CS Encoder}
For each active user $k$, $\mathbf{m}^{(k)}=[\mathbf{m}^{(k)}(1),$ $\mathbf{m}^{(k)}(2)\ldots,\mathbf{m}^{(k)}(S)]$ is the coded message output by the tree encoder. $\mathbf{m}^{(k)}(1),\mathbf{m}^{(k)}(2)\ldots,\mathbf{m}^{(k)}(S)$ are mapped in to ${i_k(1)},$ ${i_k(2)},\ldots,{i_k(S)}$, which denote the indices of the codewords in the common codebook $\mathbf{A} \in \mathbb{C}^{L_{\mathrm{p}} \times N}$, where $N=2^J$. Then the active user $k$ transmits the consecutive codewords of length $L_{\mathrm{p}}$, i.e., $\mathbf{a}_{i_k(1)}, \mathbf{a}_{i_k(2)},\ldots,\mathbf{a}_{i_k(S)}$.  

\subsection{Decoding Process}
The input to the decoder is the sum of the signals transmitted by active users plus noise after receive beamforming. The decoding process also consists of a CS decoder and tree decoder. The conventional CS decoder exploits CS techniques to recover the sub-blocks transmitted from all active users. The tree decoder forms code trees to piece these sub-blocks together to obtain the original messages.
\subsubsection{CS Decoder}
For each sub-slot s, the received signal can be expressed as
\begin{equation}
\mathbf{Y}_s = \mathbf{A}\mathbf{\Delta}\overline{\mathbf{H}}_s + \overline{\mathbf{Z}}_s = \mathbf{A}\overline{\mathbf{X}}_s + \overline{\mathbf{Z}}_s,
\end{equation}
$\overline{\mathbf{X}}_s$ is a row sparse matrix and can be recovered by CS techniques such as Approximate Message Passing (AMP) \cite{donoho2009message}.

For rich-scattering environments, an accurate and widely used statistical model for the actual channel coefficients is the Gaussian model. However, in mmWave communications, the entries $\mathbf{\overline{H}}_s$ cannot be approximated by a Gaussian distribution due to the lack of scatterers. Thus, we design a special activity detector for our considered scenario. Specifically, we approximate the unknown prior distribution with Gaussian mixture (GM) \cite{wen2014channel} and EM-GM-AMP \cite{vila2013expectation} models for activity detection and channel estimation.
The coefficients in the $i$-th column of $\overline{\mathbf{X}}_s=[\overline{\mathbf{x}}_{s,1},\overline{\mathbf{x}}_{s,2},\ldots,\overline{\mathbf{x}}_{s,N_{\mathrm{RF}}}]$ are approximated to be i.i.d with marginal pdf
\begin{equation}
  p_X(x;\rho,\bm{\omega} ,\bm{\mu} ,\bm{\nu})=(1-\rho)\delta(x)+\rho \sum \limits_{i=1}^I \omega_i \mathcal{N}(x;\mu_i,\nu_i),
\end{equation}
where $\delta(\cdot)$ is the Dirac delta, $\rho$ is the sparsity rate, and for the $k$-th GM component, $w_k$, $\mu_k$, $\nu_k$ are the weight, mean, and variance, respectively.
The sparsity of the vector is captured by the
sparsity rate $\rho$. The weights, means, and variances can be iteratively learned by the Expectation-Maximization (EM) algorithm.

For each sub-slot $s$, the CS algorithm outputs the estimation of $\mathbf{\overline{X}}_s$, i.e., $\hat{\mathbf{X}}_s$. Via maximum-ratio-combining (MRC), the activity detector ${\tilde{a}_k}(s)$ is defined as
\begin{equation}
{\tilde{a}_k}(s) = \left\{ {\begin{array}{*{20}{c}}
{1,\;  \sum\limits_{i=1}^{N_{\mathrm{RF}}} \eta_i \left| {{{\hat{x}}_{k,i}^{(s)}}} \right|  \ge {\epsilon}},\\ \\
{0,\;  \sum\limits_{i=1}^{N_{\mathrm{RF}}} \eta_i \left| {{{\hat{x}}_{k,i}^{(s)}}} \right| < {\epsilon }},
\end{array}} \right.
\end{equation}
where $\epsilon$ is a threshold, and $\eta_i$ is expressed as
\begin{equation}
    \eta_i=\frac{\left| {{{\hat{x}}_{k,i}^{(s)}}} \right|}{\sqrt{\sum\limits_{j=1}^{N_{\mathrm{RF}}}{\left| {{{\hat{x}}_{k,j}^{(s)}}} \right|}^2}}.
\end{equation}
Through the activity detector, the indices of the transmitted codewords in the common codebook $\mathbf{A}$ are obtained and collected in the set $\mathcal{K}_s$, which is written as $\mathcal{K}_s=\{k\;|\;\tilde{a}_k(s)=1,k=1,2,\ldots,N\}$. As the relationship between a sub-block and the corresponding codeword is a one-to-one mapping, if a codeword is detected, then the corresponding sub-block can be recovered automatically.
The CS Decoder finally outputs the set of the sub-blocks ${\mathcal{L}_s} = \{ \mathbf{m}_k(s)\;|\;k \in \mathcal{K}_s \}$ and the corresponding estimated channel vectors  $\mathcal{H}_s=\{ \hat {\mathbf{h}}_k^{(s)}\;|\;k \in \mathcal{K}_s\}$. Notice that the index $k$ cannot represent the identity of the active user. The information that is known at the BS is that a sub-block $\mathbf{m}_k(s)$ is transmitted, it comes from a certain user and the estimated channel gain of that user is $\hat {\mathbf{h}}_k^{(s)}$.
Besides, let $\left|\mathcal{L}_s\right|_{\mathrm{c}}=K_s$, $K_s$ means the number of the sub-blocks collected in sub-slot s. $K_s$ is usually less than $K_{\mathrm{a}}$, i.e., $K_s \le K_{\mathrm{a}}$, due to the following two reasons:
\begin{enumerate}[i)]
\item Since all users use a common codebook, the messages from different users may share some sub-blocks, which is defined as collision.
\item Due to the poor channel condition and the mistake of the CS decoder, some sub-blocks may be lost.
\end{enumerate}

\subsubsection{Tree Decoder} The traditional tree decoder in \cite{amalladinne2020coded} aims to recover the original messages transmitted from all active users by piecing together valid sequences of the sub-blocks drawn from $\mathcal{L}_1, \mathcal{L}_2,\ldots,\mathcal{L}_S$. As an initial step, the decoder fixes a sub-block in $\mathcal{L}_1$ as the root of a tree and gets the parity bits of the next sub-block by (\ref{parity}). All sub-blocks in $\mathcal{L}_2$ matching the parity bits are attached to the root. This process then moves forward. For every candidate path at stage $s$, parity bits are computed, and the matching sub-blocks in $\mathcal{L}_s$ are attached to this path, forming new branches. This continues until the last sub-slot is reached. At this point, every surviving path is output as a valid tree message.

However, the traditional tree decoder has the following two problems:
\begin{enumerate}[i)]
\item The loss of a sub-block from a particular user by the CS decoder finally leads to missed detection of the original message from that user.
\item The parity bits to be attended in every sub-block are fixed, which restricts the maximum active users that the system can serve.
\end{enumerate}

To tackle the above problems, we propose two beam-space tree decoders, which are based on hard decision and soft decision, respectively. The beam-space tree decoder with hard decision has low complexity, which is suitable to the scenario of massive connectivity. The beam-space tree decoder with soft decision considers the problem of packet loss, which can be applied to the scenario with poor channel condition. 
\begin{table}
\centering
\begin{tabular}{|c|c|}
\hline
Notation & Parameter Description \\
\hline
$\mathbf{m}_i(s)$ & The $i$-th sub-block detected in the $s$-th sub-slot \\
\hline
$c_i[l]$ & \makecell[c]{The index of the sub-block at stage $i$ \\ in the $l$-th path} \\
\hline
$\mathbf{f}_k^{(s)}$ & The beam pattern of the sub-block $\mathbf{m}_k(s)$ \\
\hline
$\mathcal{T}_s[l]$ & \makecell[c]{A set that contains the indexes of the detected  \\ sub-blocks that meet the parity constraints \\ at stage $s$ for the $l$-th path} \\
\hline 
$\mathcal{M}_s[l]$ & \makecell[c]{A set that contains the detected \\ sub-blocks that meet the parity constraints \\ at stage $s$ for the $l$-th path} \\
\hline
$\mathcal{M}^{\mathrm{hd}}_s[l]$ & \makecell[c]{A set that contains the detected   \\ sub-blocks that  meet the parity and  \\ beam pattern matching constraints \\ at stage $s$ for the $l$-th path} \\
\hline
\end{tabular}
\caption{This list contains the key parameters encountered in Section IV.}
\label{table1}
\end{table}
\section{Beam-space Tree Decoder with Hard Decision}
\begin{figure}[t]
\small
\centering
\includegraphics[width=1\linewidth]{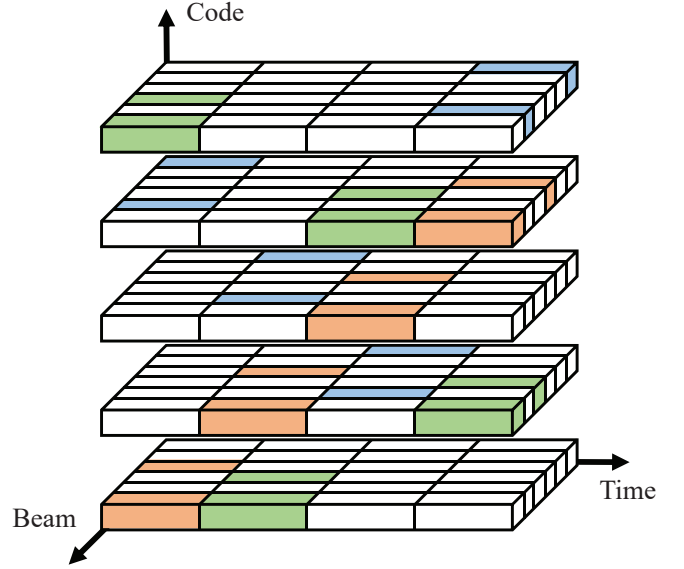}
\caption{\rm{Time}-beam-code sparse pattern.}
\label{3Dpattern}
\end{figure}

The traditional tree decoder exploits the discriminating power of parity bits to stitch the sub-blocks together to form a valid message instead of the erroneous one. At any stage of a path during the decoding process, the sub-blocks meeting the parity constraints are attached to the path. Besides the valid sub-block, other attached sub-blocks are the ones that cannot be distinguished by the parity bits. These invalid sub-blocks may finally lead to an erroneous message output by the tree decoder. Notice that in all sub-slots, the sub-blocks sent by different users are received by different beams at the BS according to the location of the users and scatterers. Therefore, the discriminating power of beams can be exploited to distinguish the invalid sub-blocks that meet the parity constraints. By leveraging the beam dimension, the decoding process can be formulated as a problem of path search in the three-dimensional space, which is shown in Fig. \ref{3Dpattern}. 
 
\begin{figure*}[!htp]
\small
\centering
\includegraphics[width=1\linewidth]{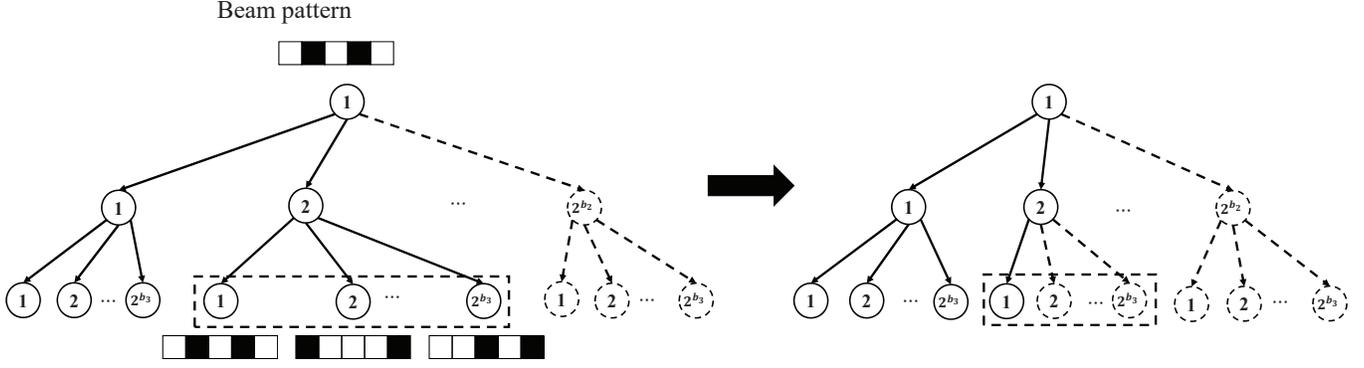}
\caption{$\mathrm{The\;pruning\;process\;of\;the\;proposed\;beam}$-$\mathrm{space\;tree\;decoder\;with\;hard\;decision.}$}
\label{pruning_hd}
\end{figure*}

To better describe the decoding process of the beam-space tree decoder with hard decision, Table \ref{table1} is given to summarize the important parameters encountered in this Section. Specifically, define the beam pattern of a sub-block as a set that contains the indices of the beams that receive the sub-block. Then different sub-blocks can be distinguished by their beam patterns.
The beam pattern $\mathbf{f}_k^{(s)}$ is written as $\mathbf{f}_k^{(s)}=[f_{k,1}^{(s)},f_{k,2}^{(s)},\ldots,f_{k,N_{\mathrm{RF}}}^{(s)}]\in \{0,1\}^{1 \times N_{\mathrm{RF}}}$.
To get accurate beam patterns, assume the gains of the active beams obey a prior known Gaussian distribution, i.e., $\mathcal{N}(\mu_1, \delta_1^2)$, where an "active" beam means that at least the signal from one user is received by the beam. And the gains of the inactive beams obey another known Gaussian distribution, i.e., $\mathcal{N}(\mu_2, \delta_2^2)$, where $\mu_1 > \mu_2$ and $\delta_1 > \delta_2$. For the gains of the inactive beams, as no signal is received or the signal experiences deep fading, $\mu_2$ is close to zero. For the gains of the active beams, $\delta_1$ is large as the signals experience random fading. Using these two prior distributions, the gains of the beams can be grouped into two classes. And for each class, the mean and variance of the samples are calculated and the prior distributions can be updated, i.e., $\mu_1 \rightarrow  \hat{\mu}_1$, $\mu_2 \rightarrow  \hat{\mu}_2$, $\delta_1 \rightarrow  \hat{\delta}_1$ and $\delta_2 \rightarrow  \hat{\delta}_2$. Then according to these updated distributions, we can give the decision rules of the beam patterns.
Specifically, $f_{k,m}^{(s)}$ is obtained by  $ \hat{h}_{k,m}^{(s)}$, which is expressed as 
\begin{equation}
f_{k,m}^{(s)}= \left\{ {
\begin{array}{*{20}{c}}
1,& \frac{\mathrm{P}_1(\left| \hat{h}_{k,m}^{(s)}\right|)}{\mathrm{P}_2(\left| \hat{h}_{k,m}^{(s)}\right|)} \ge 1,\\ \\
0,& \frac{\mathrm{P}_1(\left| \hat{h}_{k,m}^{(s)}\right|)}{\mathrm{P}_2(\left| \hat{h}_{k,m}^{(s)}\right|)} < 1.\\
\end{array}
} \right.
\label{beam_pattern}
\end{equation}
where 
$\mathrm{P}_i(\left| \hat{h}_{k,m}^{(s)}\right|)=\frac{1}{\sqrt{2 \pi}\hat{\delta}_i}e^{-\frac{\left(\left| \hat{h}_{k,m}^{(s)}\right|-\hat{\mu}_i\right)^2}{2\hat{\delta}_i^2}}$.

\begin{algorithm}[t]
\small
\caption{: Beam-Space Tree Decoder with Hard Decision}
\label{algorithm1}
\noindent{\textbf {Input}:} $\mathcal{F}_i,\mathcal{L}_i,\mathcal{K}_i,i=1,2,\ldots,S$ \\
{\textbf {Output}:} $\mathcal{L}$
\begin{algorithmic}[1]
\FOR{$k \in \mathcal{K}_1$}
\FOR{$i = 2 : S$}
\STATE For each path $l$:
\STATE Get $\mathcal{M}_i^{\rm{hd}}[l]$ according to (\ref{M_s_hd}).
\IF{$\mathcal{M}_i^{\rm{hd}}[l]$ is empty}
\STATE Delete the path $l$.
\ELSE
\STATE Attach the sub-blocks in $\mathcal{M}_i^{\rm{hd}}[l]$ to path $l$, forming new branches.
\ENDIF
\ENDFOR
\STATE Add the surviving paths rooted by $\mathbf{m}_k(1)$ to $\mathcal{L}$.
\ENDFOR
\end{algorithmic}
\end{algorithm}

For this proposed beam-space tree decoder, take the decoding process of a certain user for example. 
At the first stage, a code tree is created and a detected sub-block in the first sub-slot becomes the root of the tree and forms the first path. The root sub-block is written as $\mathbf{m}_{c_1[1]}(1)$ and its beam pattern is $\mathbf{f}_{c_1[1]}^{(1)}$. At later stages, the sub-blocks that meet the parity and the beam pattern matching constraints are kept. By meeting the parity constraints, $\mathcal{T}_s[l]$ is written as 
\begin{equation}
\mathcal{T}_s[l]=\{i\;|\;i \in \mathcal{K}_s, \mathbf{l}_{i}(s) = \sum\nolimits_{j = 1}^{s-1} {\mathbf{b}_{c_j[l]}(j){\mathbf{G}_{j,s-1}}}\}.
\end{equation}
And $\mathcal{M}_s[l]$ is obtained by
\begin{equation}
    \mathcal{M}_s[l]=\{\mathbf{m}_i(s)\;|\;i \in \mathcal{T}_s[l]\}.
\label{M_s}
\end{equation}
After beam pattern matching, only the sub-blocks in $\mathcal{M}_s^{\rm{hd}}[l]$ are survived. $\mathcal{M}_s^{\rm{hd}}[l]$ can be written as
\begin{equation}
\mathcal{M}_s^{\rm{hd}}[l]=\{\mathbf{m}_i(s)\;|\;\mathbf{f}_{c_1[l]}^{(1)} {\mathbf{f}_i^{(s)}}^{\rm{T}} \ne 0,i \in \mathcal{T}_s[l]\},
\label{M_s_hd}
\end{equation}
where $\mathcal{M}_s^{\rm{hd}}[l] \subseteq \mathcal{M}_s[l]$. Also, $\mathbf{f}_{i}^{(1)} {\mathbf{f}_{j}^{(s)}}^{\rm{T}} \ne 0$ means that the sub-blocks $\mathbf{m}_{i}(1)$ and $\mathbf{m}_{j}(s)$ are received by at least one same beam at the BS, then $\mathbf{m}_{i}(1)$ and $\mathbf{m}_{j}(s)$ have the probability to be transmitted by the same user. By beam pattern matching, the proposed beam-space tree decoder can reduce the number of surviving sub-blocks in each sub-slot, which improves the discriminating power of the decoder. A practical pruning process of this algorithm is shown in Fig. \ref{pruning_hd}. The sub-block is received by several beams at the BS, which is shown in the beam pattern. 
For every candidate path at stage $s$, there are $2^{b_s}$ candidate sub-blocks in the common codebook $\mathbf{A}$ that meet the parity constraints according to the parity bits. A part of them are inactive, while another part of them are discriminated by the beam pattern matching. As shown in Fig .4, the lines of the 2-nd and the $2^{b_3}$-th candidate sub-blocks change from solid lines to dashed lines, which means that they are deleted, as there is no overlap between their beam patterns and the beam pattern of the root sub-block at stage 1. 

Finally, the proposed beam-space tree decoder with hard decision is summarized and given in Algorithm~\ref{algorithm1}. $\mathcal{F}_s$ is a set that contains the beam patterns of the sub-blocks detected in sub-slot s, where $\mathcal{F}_s$ is expressed as $\mathcal{F}_s=\{\mathbf{f}_k^{(s)}\;|\;k \in \mathcal{K}_s\}$.

\section{Beam-space Tree Decoder with Soft Decision}
As described above, the CS decoder cannot always detect all the transmitted sub-blocks because the received signals may experience deep fading. The loss of a sub-block by the CS decoder in any sub-slot finally leads to missed detection of the original message. This is because the traditional tree decoder and the beam-space tree decoder with hard decision just stitch the sub-blocks drawn from the output of the CS decoder. At any stage, according to the parity bits, the set of candidate sub-blocks can be obtained. At stage s, the tree decoder keeps the intersection between the candidate set and $\mathcal{L}_s$. In this proposed algorithm, we keep all the candidate sub-blocks and calculate the LLR values of them by implementing the MPA algorithm, which denotes the probability of whether the sub-blocks are transmitted. Then, we define a path metric to calculate the reliability of the consecutive sub-blocks and keep some reliable paths at every stage. Even if a sub-block in a sub-slot is missed, it is possible for the path to be reliable because the path metric measures the reliability of the entire path. Therefore, the purpose of packet loss recovery is achieved.

Specifically, take a user's decoding process for example. At stage $s$ for the $l$-th path, the number of the candidate sub-blocks is $2^{b_s}$, and these sub-blocks are collected in the set $\mathcal{M}^{\prime}_s[l]$. $\mathcal{M}^{\prime}_s[l]$ is expressed as
\begin{equation}
    \mathcal{M}^{\prime}_s[l]=\{\mathbf{m}_i(s)\;|\;i \in \mathcal{T}^{\prime}_s[l]\},
\label{M_s_p}
\end{equation}
where
\begin{equation}
\mathcal{T}^{\prime}_s[l]=\{i\;|\;\mathbf{l}_{i}(s) = \sum\nolimits_{j = 1}^{s-1} {\mathbf{b}_{c_j[l]}(j){\mathbf{G}_{j,s-1}}}\}.
\end{equation}
The difference between $\mathcal{M}^{\prime}_s[l]$ and $\mathcal{M}_s[l]$ in (\ref{M_s}) is that the candidate sub-blocks in  $\mathcal{M}^{\prime}_s[l]$ are drawn according to the parity bits only, thus $\mathcal{M}_s[l] \subseteq \mathcal{M}^{\prime}_s[l]$. To reduce interference, only the received signals of those candidate sub-blocks are kept, which is denoted by $\widetilde{\mathbf{Y}}_s[l] \in \mathbb{C}^{N_{\mathrm{RF}} \times L_{\mathrm{p}}}$. $\widetilde{\mathbf{Y}}_s[l]$ is written as
\begin{equation}
\widetilde{\mathbf{Y}}_s[l]= \mathbf{Y}_s -\sum\limits_{k \in \{1,2,\ldots,N\} \backslash \mathcal{T}^{\prime}_s[l]} {\hat{\mathbf{h}}_k^{(s)}\mathbf{a}_k^{\rm{T}}},
\end{equation}
where $k \in \{1,2,\ldots,N\} \backslash \mathcal{T}^{\prime}_s[l]$ means that $k$ is in the set $\{1,2,\ldots,N\}$ instead of $\mathcal{T}^{\prime}_s[l]$. 
Then a factor graph is formed, taking the corresponding codewords as variable nodes and beam resources as resource nodes. Let $K$ and $T$ denote the number of variable nodes and resource nodes. To exploit the beam division property, the active beams in the beam pattern of the root sub-block form the resource nodes. Then the remaining received signal is defined as $\widetilde{\mathbf{Y}}^{r}_s[l] \in \mathbb{C}^{\left\| \mathbf{f}_{c_1[1]}^{(1)}\right\|_1 \times L_{\mathrm{p}}}$. A certain row of $\widetilde{\mathbf{Y}}^{r}_s[l]$ 
comes from the $j$-th row of $\widetilde{\mathbf{Y}}_s[l]$, where $j \in \{i \;|\;\mathbf{f}^{(1)}_{c_1[1],i}=1\}$.
For the sake of simplicity, define $\mathbf{y}_t$ as the received signal at the $t$-th resource, $\hat{h}_{k,t}$ as the estimated channel gain between the user transmitting the $k$-th codeword and the BS at the $t$-th resource$, \mathbf{a}_k$ as the $k$-th possible codeword and $x_k \in \{0,1\}$ as a random variable that indicates whether the codeword $\mathbf{a}_k$ is transmitted. Then $\widetilde{\mathbf{Y}}^{r}_s[l]$ is written as $\widetilde{\mathbf{Y}}^{r}_s[l]=[\mathbf{y}_1, \mathbf{y}_2,\ldots,\mathbf{y}_T]$, where $\mathbf{y}_t$ can be expressed as
\begin{equation}
{\mathbf{y}_t} = \sum\limits_{k = 1}^K {{\hat{h}_{k,t}}{x_k}{\mathbf{a}_k} + {\mathbf{z}_t} = {\hat{h}_{k,t}}{x_k}{\mathbf{a}_k} + {\mathbf{\xi} _{k,t}}},
\end{equation}
where
\begin{equation}
{\mathbf{\xi}_{k,t}} = {\mathbf{y}_t} - {\hat{h}_{k,t}}{x_k}{\mathbf{a}_k} = \sum\limits_{{k^\prime} \ne k} {{\hat{h}_{{k^\prime},t}}{x_{{k^\prime}}}{\mathbf{a}_{{k^\prime}}} + {\mathbf{z}_t}},
\end{equation}
and $\mathbf{\xi}_{k,t}=[\xi_{k,t,1},\xi_{k,t,2},\ldots,\xi_{k,t,L_{p}}]$. To reduce the computational complexity, we resort to Gaussian Approximation (GA) as in \cite{idma}, and approximate ${\xi _{k,t,j}}$ as a complex Gaussian-distributed random variable with mean $\mu _{{\xi _{k,t,j}}}$ and variance $\delta _{{\xi _{k,t,j}}}^2$, i..e., ${\xi _{k,t,j}} \sim \mathcal{CN}({\mu _{{\xi _{k,t,j}}}},\delta _{{\xi _{k,t,j}}}^2)$. ${\mu _{{\xi _{k,t,j}}}}$ and $\delta _{{\xi _{k,t,j}}}^2$ can be expressed as
\begin{figure*}[!htp]
\small
\centering
\includegraphics[width=1\linewidth]{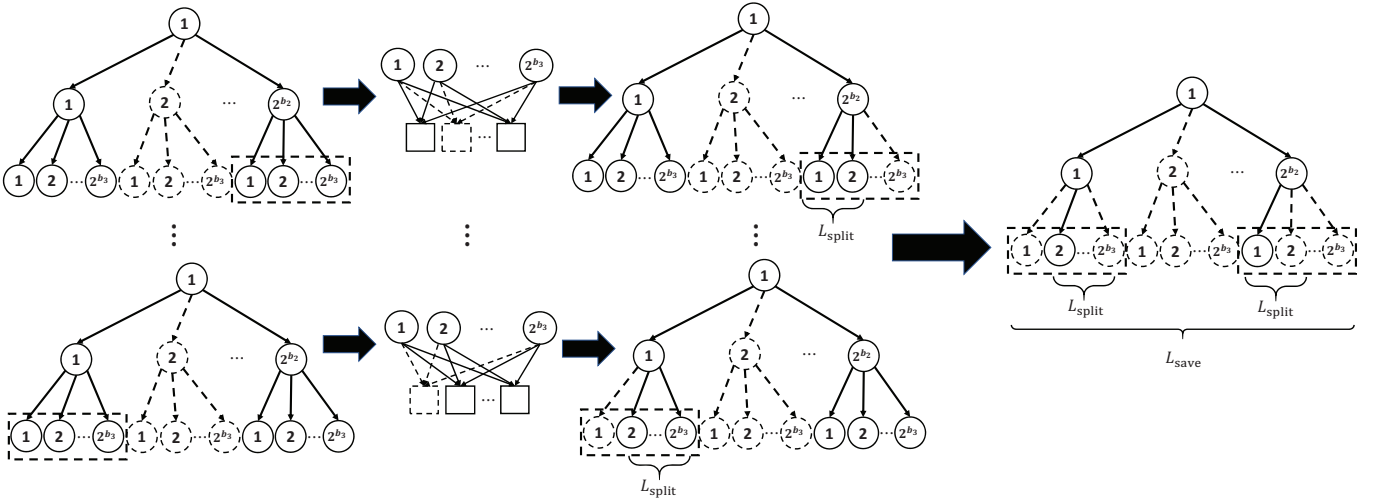}
\caption{$\mathrm{The\;pruning\;process\;of\;the\;proposed\;beam}$-$\mathrm{space\;tree\;decoder\;with\;soft\;decision.}$}
\label{pruning_sd}
\end{figure*}
\begin{equation}
\begin{aligned}
{\mu _{{\xi _{k,t,j}}}} &= \sum\limits_{{k^\prime} \ne k} {{\hat{h}_{{k^\prime},t}}{a_{{k^\prime},j}}{P_{k^\prime \to t}}},\\
\delta _{{\xi _{k,t,j}}}^2 &= \sum\limits_{k^\prime \ne k} {{{| {{\hat{h}_{k^\prime,t}}{a_{k^\prime,j}}} |}^2}{P_{k^\prime \to t}}(1 - {P_{k^\prime \to t}}) + \delta _{{z_{t,j}}}^2},
\end{aligned}
\end{equation}
respectively, where
\begin{equation}
P_{k^\prime \to t}=\frac{\exp(L_{k^\prime \to t})}{1+\exp(L_{k^\prime \to t})}.
\end{equation}
$L_{k \to t}$ is the log likelihood ratio (LLR) delivered from the $k$-th variable node to the $t$-th resource node. Also, $L_{t \to k}$ denotes the LLR delivered from the $t$-th resource node to the $k$-th variable node, which is written as
\begin{equation}
{L_{t \to k}} = \ln \frac{{p({\mathbf{y}_t}|{x_k} = 1)}}{{p({\mathbf{y}_t}|{x_k} = 0)}},
\end{equation}
where
\begin{equation}
\begin{aligned}
&p({\mathbf{y}_t}|{x_k} = 1) \\ &= \prod\limits_{j = 1}^{{L_{\mathrm{p}}}} {\frac{1}{{\sqrt {2\pi \delta _{{\xi _{k,t,j}}}^2} }}\exp \left( { - \frac{{{{| {{y_{t,j}} - {\hat{h}_{k,t}}{a_{k,j}} - {\mu _{{\xi _{k,t,j}}}}} |}^2}}}{{2\delta _{{\xi _{k,t,j}}}^2}}} \right)},
\end{aligned}
\end{equation}
\begin{equation}
p({\mathbf{y}_t}|{x_k} = 0) =\prod\limits_{j = 1}^{{L_{\mathrm{p}}}} {\frac{1}{{\sqrt {2\pi \delta _{{\xi _{k,t,j}}}^2} }}\exp \left( { - \frac{{{{| {{y_{t,j}} - {\mu _{{\xi _{k,t,j}}}}} |}^2}}}{{2\delta _{{\xi _{k,t,j}}}^2}}} \right)}.
\end{equation}
Besides, $L_{k \to t}$ is given as
\begin{equation}
{L_{k \to t}} = \sum\limits_{t^{\prime} \ne t} {{L_{t^{\prime} \to k}}}.
\end{equation}
Finally, $L_{k}$ can be expressed as
\begin{equation}
{L_{k}} = \sum\limits_{t=1}^T {{L_{t \to k}}}.
\end{equation}
Denote $l^{(n)}$ as the new paths that are split from the $l$-th path. By implementing MPA, from path $l$ at stage s, we can obtain the LLRs of the candidate sub-blocks in $\mathcal{M}^{\prime}_s[l]$, which are written as $L_s[l^{(n)}],n=1,2,\ldots,2^{b_s}$. Learning from the way of list decoding \cite{tal2015list}, we define a path metric to calculate the reliability of the new branches from path $l$ at stage $s$. Specifically, the PM of the new branch $l^{(n)}$ is written as
\begin{equation}
\begin{aligned}
\mathrm{PM}_s[l^{(n)}] &= \sum\limits_{i = 1}^{s-1} {\ln (1 + {e^{ - {L_i[l]}}})+\ln (1 + {e^{ - {L_s[l^{(n)}]}}})} \\ &= \mathrm{PM}_{s-1}[l]+\ln (1 + {e^{ - {L_s[l^{(n)}]}}}),
\end{aligned}
\label{PM}
\end{equation}
where $L_i[l]$ denotes the LLR of the sub-block at stage $i$ in the path $l$. The decoder calculates the PM of all the branches from every candidate path and keep some reliable paths at every stage. And at the last stage, the decoder outputs the most reliable path as the recovered message.

However, this scheme is not suitable in the case that collision occurs. As mentioned above, active users select codewords from a common codebook $\mathbf{A}$. Even if the dimension of $\mathbf{A}$ is large, collisions may still occur. If the traditional tree decoder fixes a sub-block transmitted by two active users at the first stage, then the decoder finally outputs two valid tree messages.
According to \cite{fengler2020pilot}, we give $\mathbf{E}\{C_{k,s}\}$ as the average number of collisions of $k$ users on consistent s sub-blocks started from the first one, which is written as
\begin{equation}
    \mathbf{E}\{C_{k,s}\}=\frac{\binom{K_{\mathrm{a}}}{k}}{{(N{\prod\limits_{i = 2}^s {{2^{{b_s}}}}})}^{k-1}}.
\end{equation}
The collision of more than two users is ignored because the number is usually much smaller than 1. As $s$ grows, the collision can be ignored when $k=2$. In other words, it is impossible for the valid messages of the collision users to be the new branches of the same candidate path. However, when the depth of a code tree grows, the LLR of a sub-block has less impact on the PM of the entire path. Therefore, the remaining paths at stage s may all come from the new branches of the most reliable path at stage $s-1$, leading to missed detection. Actually, there is no need to keep all new branches from a candidate path because the valid messages of collision users come from different candidate paths.
Denote $L_{\mathrm{split}}$ as the number of splitting paths, which means that only $L_{\mathrm{split}}$ new branches from a candidate path are kept according to the PM. Then for the current stage, keep $L_{\mathrm{save}} $ most reliable paths, where $L_{\mathrm{save}} > L_{\mathrm{split}}$. This pruning process is shown in Fig. \ref{pruning_sd}. 
The number of candidate sub-blocks at stage $s$ is $2^{b_s}$, which is equivalent to the process in Fig. 4. At stage $s$, for the new branches of a path, a factor graph is created to implement the MPA algorithm and give the candidate sub-blocks LLRs. Then the most reliable $L_{\mathrm{split}}$ paths are kept, and others are deleted, which is shown in Fig. \ref{pruning_sd} that the lines of those deleted sub-blocks change from solid to dashed. Finally at stage $s$, for the $2^{b_{s-1}}L_{\mathrm{split}}$ paths, the most reliable $L_{\mathrm{save}}$ paths are kept and others are deleted.
$L_{\mathrm{split}}$ is chosen according to the trade-off between the computational complexity and decoding performance of the decoder. 

\begin{algorithm}[t]
\small
\caption{: Beam-Space Tree Decoder with Soft Decision}
\label{algorithm2}
\noindent{\textbf {Input}:} $\mathcal{F}_i,\mathcal{L}_i,\mathcal{K}_i,i=1,2,\ldots,S$ \\
{\textbf {Output}:} $\mathcal{L}$
\begin{algorithmic}[1]
\FOR{$k \in \mathcal{K}_1$}
\FOR{$i = 2 : S$}
\IF {$i \le S^{\prime}$}
\STATE For each path $l$:
\STATE Get $\mathcal{M}^{\prime}_i[l]$ according to (\ref{M_s_p}).
\STATE Form a factor graph and implement MPA algorithm to get LLR values of the sub-blocks in $\mathcal{M}^{\prime}_i[l]$.
\STATE Calculate $\mathrm{PM}_i[l^{(n)}]$ by (\ref{PM}), keep the $L_{\mathrm{split}}$ most reliable paths and delete others.
\STATE For all the new paths: 
\STATE Keep the $L_{\mathrm{save}}$ most reliable paths and delete others.
\ELSE
\STATE For each path $l$:
\STATE Get $\mathcal{M}^{\mathrm{hd}}_i[l]$ according to (\ref{M_s_hd}).
\IF { $\mathcal{M}^{\mathrm{hd}}_i[l]$ is empty}
\STATE Delete the path $l$.
\ELSE
\STATE Attach the sub-blocks in $\mathcal{M}^{\mathrm{hd}}_i[l]$ to path $l$, forming new branches.
\ENDIF
\ENDIF
\ENDFOR
\STATE For every pair $\left\langle\mathbf{u}_m,\mathbf{u}_n\right\rangle$ in the surviving paths:
\STATE Get $\mathrm{P_s}(\mathbf{u}_m,\mathbf{u}_n)$ according to (\ref{P_s}).
\IF {$\mathrm{P_s}(\mathbf{u}_m,\mathbf{u}_n) > \tau$}
\STATE Keep the more reliable one according to (\ref{PM}) and delete the other.
\ENDIF
\STATE Add the remaining paths to $\mathcal{L}$.
\ENDFOR
\end{algorithmic}
\end{algorithm}

At the last stage, the number of messages output by the decoder cannot be determined because whether a collision occurs is unknown to the decoder. Notice that the traditional tree decoder outputs all the paths meeting the parity constraints as valid messages. When a collision occurs, the traditional tree decoder outputs several paths as the recovered messages. Besides, more parity bits are pushed towards later stages to reduce the probability of error according to \cite{amalladinne2020coded}. Thus, we exploit the discriminating power of the parity bits at later stages to output the results of the beam-space tree decoder with soft decision. To summarize, list decoding is implemented at the former $S^{\prime}$ stages to keep the missed sub-blocks, and the sub-blocks that are full of parity bits are exploited to prune the erroneous paths at the latter $S-S^{\prime}$ stages.

However, due to the loss packet recovery, the missed detection rate decreases while the false alarm rate increases. The reason is that some undetected sub-blocks may be kept in the decoding process. Some of them are not transmitted actually, which may lead to false alarm. As the valid messages from collision users are not similar with each other, the invalid messages can be discriminated. Specifically, define a similarity metric $\mathrm{P_s}({\mathbf{u}_i},{\mathbf{u}_j})$, which is denoted by
\begin{equation}
    \mathrm{P_s}({\mathbf{u}_i},{\mathbf{u}_j}) = \frac{{\sum\limits_{s = 1}^S {\mathbb{I}({\mathbf{b}_i}(s) = {\mathbf{b}_j}(s))} }}{S},
\label{P_s}
\end{equation}
where $\mathbb{I}(\cdot)$ is the indicator function, $\mathbf{u}_i$ and $\mathbf{u}_j$ are two messages output by a code tree. Calculate every pair of the outputs from a code tree, if $\mathrm{P_s}({\mathbf{u}_i}$, ${\mathbf{u}_j}) > \tau$, then keep the more reliable one according to the PM, otherwise keep both, where $\tau$ is a threshold. By leveraging the similarity metric, the invalid messages meeting the parity constraints are deleted.
The beam-space tree decoder with soft decision is summarized and given in Algorithm~\ref{algorithm2}.

\section{Performance Analysis}
The performance of our proposed URA system is connected with the reliability of CS techniques in each sub-slot and the efficiency of message stitching across different sub-slots. In the remainder of this section, we ignore the collision that different active users share a sub-block in the first sub-slot.

Take the decoding process of user $k$ for example. Let $p_{\rm{cs}}$ be the probability that at least one sub-block of user $k$ is not output by the CS decoder, $p_{\rm{tree}}$ be the probability of error, $p_{\rm{tree}}^{\mathrm{md}}$ be the probability of missed detection, and $p_{\rm{tree}}^{\mathrm{fa}}$ be the probability of false alarm. $p_{\rm{tree}}$ is written as
\begin{equation}
\begin{aligned}
    p_{\rm{tree}}&=p_{\rm{tree}}^{\mathrm{md}}+p_{\rm{tree}}^{\mathrm{fa}}\\&=p_{\rm{cs}}p_{\rm{tree}|cs}^{\mathrm{md}}+p_{\rm{\overline{cs}}}p_{\rm{tree}|\overline{cs}}^{\mathrm{md}}+p_{\rm{tree}}^{\mathrm{fa}},
\end{aligned}
\end{equation}
where $p_{\overline{cs}}=1-p_{\rm{cs}}$, which denotes the probability that the CS decoder is error-free. $p_{\rm{tree}|\overline{cs}}^{\mathrm{md}}$ denotes the probability that the message of user $k$ is not output by the tree decoder in the case that the CS decoder is error-free. 
\subsection{Tree Code Analysis}
In this subsection, we analyze the error probability of all the tree decoders. Denote $p_{\rm{tree}}$, $\tilde{p}_{\rm{tree}}$, $\hat{p}_{\rm{tree}}$ as the error probability of the traditional tree decoder, the beam-space tree decoder with hard decision and the beam-space tree decoder with soft decision.

For the traditional tree decoder and the beam-space tree decoder with hard decision, it is obvious that $p_{\rm{tree}|cs}^{\mathrm{md}}=1,\tilde{p}_{\rm{tree}|cs}^{\mathrm{md}}=1$ because these two decoders cannot deal with the problem of packet loss. In contrast, $p_{\rm{tree}|\overline{cs}}=0,\tilde{p}_{\rm{tree}|\overline{cs}}=0$ because there is no missed detection when the CS decoder is error-free. Thus, we use $p_{\rm{tree}}^{\mathrm{fa}},\tilde{p}_{\rm{tree}}^{\mathrm{fa}}$ to analyze the performance of the two decoders, respectively.

 Let $L^{(j)}$, $\tilde{L}^{(j)}$ be the number of erroneous paths of the traditional tree decoder and the beam-space tree decoder with hard decision at stage $j$, respectively. 
 Define $p_{\rm{match}}$ as the probability of the event that the intersection between the beam patterns of two sub-blocks are not empty. Assume the signal transmitted from an active user is received by $N_{\mathrm{b}}$ beams at the BS. According to (\ref{M_s_hd}), the beam pattern matching is implemented by $\mathbf{f}_{i}^{(1)} {\mathbf{f}_{j}^{(s)}}^{\rm{T}} \ne 0$. Then $p_{\rm{match}}$ is given as
\begin{equation}
    p_{\rm{match}}=\frac{\sum\limits_{i=1}^{N_{\mathrm{b}}}{\binom{N_{\mathrm{b}}}{i}\binom{N_{\rm{RF}}-N_{\mathrm{b}}}{N_{\mathrm{b}}-i}}}{\binom{N_{\rm{RF}}}{N_{\mathrm{b}}}}=1-\frac{\binom{N_{\rm{RF}}-N_{\mathrm{b}}}{N_{\mathrm{b}}}}{\binom{N_{\rm{RF}}}{N_{\mathrm{b}}}}.
\label{p_match}
\end{equation}
$p_{\rm{match}} \ll 1$ because of the lack of scatterers in the mmWave communication system, i.e., $N_{\mathrm{b}} \ll N_{\mathrm{RF}}$.

For the beam-space tree decoder with hard decision, we give the expected values of $\tilde{L}^{(j)}$ in Theorem \ref{theorem1}.
\begin{theorem}
The expected values of $\tilde{L}^{(j)}$, which is denoted by $\mathbb{E}[\tilde{L}^{(j)}]$, can be expressed as
\begin{equation}
\mathbb{E}[\tilde{L}^{(j)}] = \sum\limits_{q = 2}^j {p_{\rm{match}}}^{j-q+1}{{K^{j - q}}(K - 1)\prod\limits_{s = q}^j {{2^{ - {l_s}}}} },
\label{E_L}
\end{equation}
where $K=K_{\mathrm{a}}$, $j =2,\ldots,S$. 
\label{theorem1}
\end{theorem}
\begin{IEEEproof}
According to \cite{amalladinne2020coded}, for $j \ge 3$, $\mathbb{E}[{L}^{(j)}]$ is written as 
\begin{equation}
    \mathbb{E}[{L}^{(j)}]=2^{-l_j}\mathbb{E}[{L}^{(j-1)}]+2^{-l_j}(K-1).
\end{equation}
For $j=2$, $\mathbb{E}[{L}^{(2)}]$ is given as
\begin{equation}
\mathbb{E}[{L}^{(2)}]=(K-1)2^{-l_2}.
\end{equation}
By implementing beam pattern matching, the expected erroneous paths $\mathbb{E}[\tilde{L}^{(j)}]$ at every stage are reduced. For $j=2$, $\mathbb{E}[\tilde{L}^{(2)}]$ can be expressed as
\begin{equation}
\mathbb{E}[\tilde{L}^{(2)}]=p_{\mathrm{match}}(K-1)2^{-l_2}.
\label{E_L_iter}
\end{equation}
For $j \ge 3$, $\mathbb{E}[{L}^{(j)}]$ is written as
\begin{equation}
    \mathbb{E}[\tilde{L}^{(j)}]=2^{-l_j}p_{\mathrm{match}}\mathbb{E}[\tilde{L}^{(j-1)}]+2^{-l_j}p_{\mathrm{match}}(K-1).
\end{equation}
Using (\ref{E_L_iter}) as initial condition, $\mathbb{E}[\tilde{L}^{(j)}]$ is rewritten as (\ref{E_L}).
\end{IEEEproof}

Based on Theorem \ref{theorem1}, we give the upper bound of the false alarm rate of the beam-space tree decoder with hard decision in Corollary \ref{corollary1}.
\begin{corollary}
$\tilde{p}_{\mathrm{tree}}^{\mathrm{fa}}$ is bounded by
\begin{equation}
    \tilde{p}_{\mathrm{tree}}^{\mathrm{fa}} \le \mathbb{E}[\tilde{L}^{(S)}]. 
\end{equation}
\label{corollary1}
\end{corollary}
\begin{IEEEproof}
$\tilde{p}_{\mathrm{tree}}^{\mathrm{fa}}$ is the probability of false alarm, which means that the number of erroneous paths at the last stage is at least one. Therefore, $\tilde{p}_{\mathrm{tree}}^{\mathrm{fa}}$ is written as 
\begin{equation}
\tilde{p}_{\mathrm{tree}}^{\mathrm{fa}}= \mathrm{Pr}(\tilde{L}^{(S)} \ge 1) \le \mathbb{E}[\tilde{L}^{(S)}].
\label{p_tree}
\end{equation}
The inequality (\ref{p_tree}) is obtained by the application of Markov inequality.
\end{IEEEproof}

For the beam-space tree decoder with soft decision, $\hat{p}_{\rm{tree}|cs}^{\mathrm{md}}<1$ because the decoder considers the problem of packet loss. $\hat{p}_{\rm{tree}|\overline{cs}}^{\mathrm{md}}>0$ because the decoder keeps limit reliable paths at every stage. As $L_{\rm{save}}$ grows, $\hat{p}_{\rm{tree}|\overline{cs}}^{\mathrm{md}}$ decreases. Besides, notice that the decoder outputs the most reliable path at the last stage. If the recovered message is invalid, then missed detection and false alarm occur simultaneously, i.e., $\hat{p}_{\rm{tree}}^{\mathrm{md}}=\hat{p}_{\rm{tree}}^{\mathrm{fa}}$.
\subsection{CS Analysis}
A fundamental limitation of compressed sensing is that the required signal dimension $L_{\mathrm{p}}$ to reliably identify a subset of $K_{\mathrm{a}}$ transmitted codewords among a set consisting of $N$ codewords in the common codebook scales as $L_{\mathrm{p}}=\mathcal{O}(K_{\mathrm{a}}\mathrm{log}\frac{N}{K_{\mathrm{a}}})$. $L_{\mathrm{p}}$ is almost linearly with $K_{\mathrm{a}}$.
$K_{\mathrm{a}}$ is bounded by $K_{\mathrm{a}}=\mathcal{O}(L_{\mathrm{p}}/\mathrm{log}\frac{N}{L_{\mathrm{p}}})$.

In our scenario, although $K_{\mathrm{a}}$ active users transmit codewords simultaneously, only a small number of codewords is received by a certain beam at the BS. The average number of codewords received by a beam is given as
\begin{equation}
    \overline{K}_{\mathrm{a}}=\frac{\binom{N_{\rm{RF}}-1}{N_{\mathrm{b}}-1}}{\binom{N_{\rm{RF}}}{N_{\mathrm{b}}}}K_{\mathrm{a}}=\frac{N_{\mathrm{b}}K_{\mathrm{a}}}{N_{\mathrm{RF}}}.
\end{equation}
Denote $p_{\mathrm{b}}=\frac{N_{\mathrm{b}}}{N_{\mathrm{RF}}}$, then $K_{\mathrm{a}}$ is bounded by
\begin{equation}
K_{\mathrm{a}}=\mathcal{O} \left( \frac{L_{\mathrm{p}}}{p_{\mathrm{b}}\mathrm{log}\frac{N}{L_{\mathrm{p}}}} \right). 
\end{equation}
Similarly, $L_{\mathrm{p}}$ is bounded by 
\begin{equation}
L_{\mathrm{p}}=\mathcal{O}\left(\overline{K}_{\mathrm{a}}\mathrm{log}\frac{N}{\overline{K}_{\mathrm{a}}}\right)=\mathcal{O}\left(p_{\mathrm{b}}K_{\mathrm{a}}\mathrm{log}\frac{N}{p_{\mathrm{b}}K_{\mathrm{a}}}\right).
\label{L_p}
\end{equation}

Considering the limitation of both the CS decoder and the tree decoder, we give the upper bound of the number of active users $K_{\mathrm{a}}$ in Lemma \ref{lemma1}. 
\begin{lemma}
In our scenario, the number of active users, i.e., $K_{\mathrm{a}}$, is bounded by
\begin{equation}
    K_{\mathrm{a}} \le \min \left( {c_1\frac{L_{\mathrm{p}}}{p_{\mathrm{b}}\mathrm{log}\frac{N}{L_{\mathrm{p}}}}, \frac{2^{J(1-R_{\mathrm{tree}})+1}}{p_{\mathrm{b}}}} \right),
\end{equation}
where $c_1$ is a constant.
\label{lemma1}
\end{lemma}
\begin{IEEEproof}
Denote the average number of sub-blocks detected in a certain beam as $K_{\mathrm{v}}$, where $K_{\mathrm{v}}=p_{\mathrm{b}}K_{\mathrm{a}}$. According to the limitation of the tree decoder given in \cite{fengler2019massive}, $K_{\mathrm{v}}$ is bounded by $K_{\mathrm{v}} \le 2^{J(1-R_{\mathrm{tree}})+1}$. Substituting $K_{\mathrm{v}}$ by $K_{\mathrm{a}}$ in the inequality, the bound of $K_{\mathrm{a}}$ is obtained, i.e., $K_{\mathrm{a}} \le \frac{2^{J(1-R_{\mathrm{tree}})+1}}{p_{\mathrm{b}}}$. 
\end{IEEEproof}

\subsection{Asymptotic Analysis}
In this subsection, we study the proposed algorithms in the context of large settings. The asymptotic analysis in \cite{amalladinne2020coded} shows that the probability of false alarm goes to zero in the logarithmic regime with constant code rate $R_{\mathrm{tree}}$. In this subsection, we analyze the impact of the beam division property on system performance in the asymptotic regime while fixing the number of parity bits. 
\begin{lemma}
Fix $N_{\mathrm{RF}}=\alpha K_{\mathrm{a}}$, for some $\alpha < 1$ and consider the number of active users $K_{\mathrm{a}} \to \infty$. For the CS decoder, the required signal dimension $L_{\mathrm{p}}$ is given as
\begin{equation}
 L_{\mathrm{p}} = \mathcal{O}\left(\frac{1}{c_2}\mathrm{log}(c_2 N)\right),
\end{equation}
\label{lemma2}
where $c_2$ is a constant.
\end{lemma}
\begin{IEEEproof}
According to (\ref{L_p}), $L_{\mathrm{p}}$ is rewritten as
\begin{equation}
\begin{aligned}
    L_{\mathrm{p}}&=\mathcal{O}\left(K_{\mathrm{a}}p_{\mathrm{b}}\mathrm{log}\frac{N}{K_{\mathrm{a}}p_{\mathrm{b}}}\right) 
    =\mathcal{O}\left(\frac{N_{\mathrm{b}}K_{\mathrm{a}}}{N_{\mathrm{RF}}}\mathrm{log}\frac{N}{\frac{N_{\mathrm{b}}K_{\mathrm{a}}}{N_{\mathrm{RF}}}}\right) \\
    &= \mathcal{O}\left(\frac{1}{c_2}\mathrm{log}(c_2 N)\right).
\end{aligned}
\end{equation}
This completes the proof.
\end{IEEEproof}

Lemma \ref{lemma2} shows that as the number of RF chains and active users increases together while keeping a fixed ratio, the number of active users $K_{\mathrm{a}}$ plays no role in the required signal dimension $L_{\mathrm{p}}$. This means that the decoding performance of the CS decoder is guaranteed by exploiting the beam division property while keeping the dimension of the common codebook $\mathbf{A}$.
\begin{theorem}
Fix $N_{\mathrm{RF}}=\alpha {K_{\mathrm{a}}}$, $J=\mathrm{log}\frac{K_{\mathrm{a}}}{\varepsilon}$ for some $0<\alpha<1$, $\varepsilon \ll 1$, consider the number of active users $K_{\mathrm{a}} \to \infty$ and the number of paths $L_{\mathrm{save}} \to \infty$. The decoding performance of the beam-space tree decoder with hard decision and the beam-space tree decoder with soft decision, i.e., $\tilde{p}^{\mathrm{fa}}_{\mathrm{tree}}$ and $\hat{p}^{\mathrm{fa}}_{\mathrm{tree}}$, is close to zero.
\label{theorem2}
\end{theorem}
\begin{IEEEproof}
For the beam-space tree decoder with hard decision, according to (\ref{p_match}), $p_{\mathrm{match}}K_{\mathrm{a}} = (1-\frac{\binom{N_{\mathrm{RF}}-N_{\mathrm{b}}}{N_{\mathrm{b}}}}{\binom{N_{\mathrm{RF}}}{N_{\mathrm{b}}}})$ $\frac{1}{\alpha}N_{\mathrm{RF}} \to \frac{N_{\mathrm{b}}^2}{\alpha}$. For the sake of similarity,  assume $2^{-l_s} = c_3 \ll 1$ for s = 2,3,\ldots,S. Then $\mathbb{E}[\tilde{L}^{S}]$ is approximated as
\begin{equation}
\begin{aligned}
    \mathbb{E}[\tilde{L}^{S}] &\to \sum\limits_{q = 2}^S {(\frac{N_{\mathrm{b}}^2}{\alpha})}^{S-q+1}c_3^{S-q+1} = \sum\limits_{q = 2}^S {(\frac{c_3N_{\mathrm{b}}^2}{\alpha})}^{S-q+1} \\
    &= \frac{c_4(1-c_4^{S-1})}{1-c_4} \approx c_4,
\end{aligned}
\end{equation}
where $c_4$ is a constant.

According to Corollary \ref{corollary1}, $\tilde{p}_{\mathrm{tree}}^{\mathrm{fa}}$ is rewritten as
\begin{equation}
    \tilde{p}_{\mathrm{tree}}^{\mathrm{fa}} \le \mathbb{E}[\tilde{L}^{(S)}] \approx c_4 \ll 1.
\end{equation}

For the beam-space tree decoder with soft decision, let the number of paths $L_{\mathrm{save}} \to \infty$. At every stage, although $2^{b_s} \to \infty$ sub-blocks are drawn by the checking relationship, most of the new paths are not reliable. Only $p_{\mathrm{match}}K_{\mathrm{a}}+1 \to \frac{N_{\mathrm{b}}^2}{\alpha}+1$ codewords are needed to be verified. Without the process of pruning, $(\frac{N_{\mathrm{b}}^2}{\alpha}+1)^{S-1}$ possible paths are kept at the last stage. As $L_{\mathrm{save}} \to \infty$, the valid path is reliable and kept. At the last stage, the valid path has a high probability of being the most reliable path according to the PM and being output by the decoder as a result. 
\end{IEEEproof}

It can be seen from Theorem \ref{theorem2} that in the regime, the decoding performance of the beam-space tree decoder with hard decision and the beam-space tree decoder with soft decision is guaranteed by exploiting the beam division property.

\subsection{Computational Complexity Analysis}
We denote $C_{\rm{tree}},\tilde{C}_{\rm{tree}},\hat{C}_{\rm{tree}}$ as the computational complexity of the traditional tree decoder, the beam-space tree decoder with hard decision and the beam-space tree decoder with soft decision. For the sake of simplicity, let $K$ be the candidate sub-blocks in every sub-slot and ignore the collision. According to \cite{amalladinne2020coded}, the expected computational computational complexity of the traditional tree decoder, i.e., $\mathbb{E}[C_{\mathrm{tree}}]$, is given as
\begin{equation}
    \mathbb{E}[C_{\mathrm{tree}}]=(S-1)K+\sum\limits_{j=2}^{S-1}{\mathbb{E}[L^{(j)}]}K.
\end{equation}

The computational complexity of the beam-space tree decoder with hard decision, i.e., $\mathbb{E}[\tilde{C}_{\mathrm{tree}}]$, is written as
\begin{equation}
    \mathbb{E}[\tilde{C}_{\mathrm{tree}}]=(S-1)K+\sum\limits_{j=2}^{S-1}{\mathbb{E}[\tilde{L}^{(j)}]K}.
\end{equation}
Notice that $\mathbb{E}[\tilde{C}_{\mathrm{tree}}] \ll \mathbb{E}[C_{\mathrm{tree}}]$ because $\mathbb{E}[\tilde{L}^{(j)}] \ll \mathbb{E}[{L}^{(j)}]$. Therefore, the beam-space tree decoder with hard decision has low computational complexity.

For the beam-space tree decoder with soft decision, the computational complexity $\hat{C}_{\mathrm{tree}}$ is given as
\begin{equation}
    \hat{C}_{\mathrm{tree}}=N_{b}2^{b_{2}}I_{\mathrm{max}}+\sum\limits_{j=3}^{S}{L_{\mathrm{save}}N_{b}2^{b_{j}}I_{\mathrm{max}}},
\end{equation}
where $2^{b_j}$ and $N_{\mathrm{b}}$ are the number of variable nodes and resource nodes of the factor graph, $I_{\mathrm{max}}$ is the max number of iterations of MPA.

\section{Numerical Results}
In this section, we investigate the performance of our proposed scheme in terms of error probability. We consider a simulation scenario where the BS employs a ULA antenna array with $N_{\mathrm{r}}=256$, and the RF chains are set to $N_{\mathrm{RF}}=16$. The multipath channel consists of $P_c=3$ clusters each of which containing $Q_p=10$ sub-paths. It is assumed that the location of users obeys a two-dimensional Poisson distribution. The transmit power is fixed as $P=20$ dBm for all users. For the tree code scheme, $B=94$ bits of information are split into $S=32$ sub-blocks with length $J=10$ bits. Data profile and parity profile are given as $[10,3,\ldots,3,0,0,0]$ and $[0,7,\ldots,7,10,10,10]$, respectively. For the beam-space tree decoder with soft decision, $L_{\rm{save}}=24, L_{\rm{split}}=8$. As a reference, we compare the proposed beam-space tree decoders with the traditional tree decoder \cite{amalladinne2020coded} and the CB-CS algorithm \cite{fengler2021non}. For the sake of simplicity, in the following figures, TD means tree decoder, HD means hard decision, and SD means soft decision.

\begin{figure}[!htp]
    \small
    \centering
    \includegraphics[width=1\linewidth]{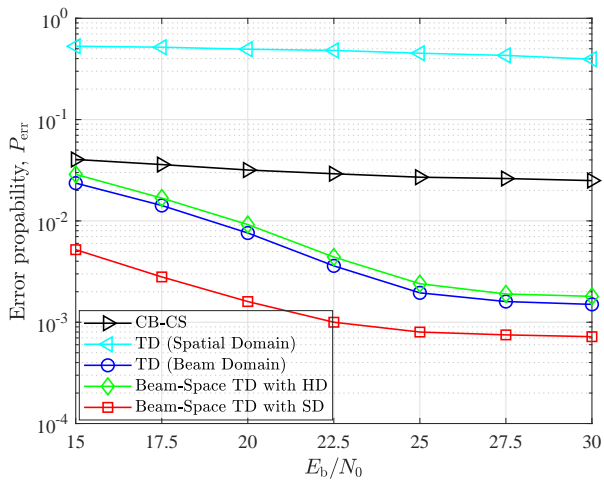}
    \caption{\rm{The} error probability of different URA schemes versus $E_{\mathrm{b}}/N_0$ when $K_{\mathrm{a}}=50$.}
    \label{ebn0_perr}
    \end{figure}
Fig. \ref{ebn0_perr} depicts the error probability of different URA schemes versus $E_{\mathrm{b}}/N_0$ when $K_{\mathrm{a}}=50$. The difference between the tree decoder (beam domain) and the tree decoder (spatial domain) is whether the tree decoding process is performed after receive beamforming. Comparing these two algorithms, we can see that the decoding performance is significantly improved after receive beamforming because the detection performance of CS techniques is improved by exploiting the sparsity of beam domain channel and the beamforming gain. 
Besides, at the BS the signals are received after beamforming, which means that the dimension of the received signals is determined by the number of RF chains instead of the number of antennas. Therefore, the high dimension of antennas can not be exploited by the CB-CS algorithm in mmWave scenarios.
In addition, there is a gap between the beam-space tree decoder with hard decision and the traditional tree decoder (beam domain). The reason is that getting accurate beam patterns of the sub-blocks, which means that whether the signal is received by every beam at the BS should be accurately determined, is harder than the activity detection of these sub-blocks. Therefore, the beam patterns of the sub-blocks obtained by (\ref{beam_pattern}) may not always be accurate, resulting in a decrease in the decoding performance. Besides, it is observed that for the considered $E_{\mathrm{b}}/N_0$, the beam-space tree decoder with soft decision achieves the best performance. Such advantages come from the fact that this algorithm considers the problem of packet loss.

\begin{figure}[!htp]
    \small
    \centering
    \includegraphics[width=1\linewidth]{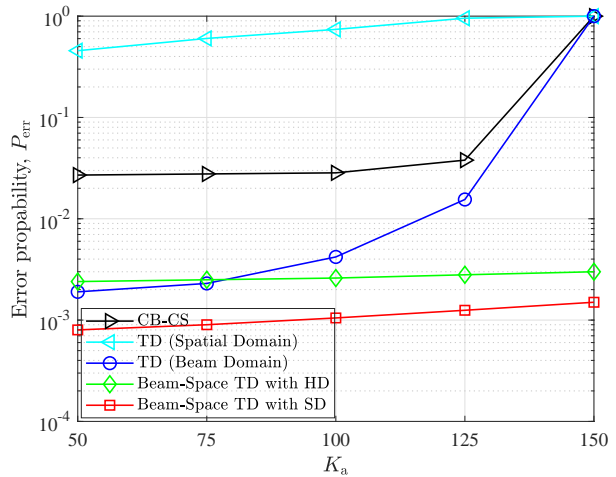}
    \caption{\rm{The} error probability of different URA schemes versus $K_{\mathrm{a}}$ when $E_{\mathrm{b}}/N_0=25\;\rm{dB}$.}
    \label{ka_perr_150}
    \end{figure}
Fig. \ref{ka_perr_150} plots the decoding performance with different active users when $E_{\mathrm{b}}/N_0=25$ dB.  
Initially, in the regime with a few number of active users, the error probability of all algorithms increases slowly. However, when the number of active users continues to increase, the error probability of the tree decoder increases sharply to 1, which means that the tree decoder cannot discriminate and stitch the valid sub-blocks when the number of active users is large. Actually, allocating more parity bits can help the traditional tree decoder accommodate more active users while the computational complexity of the CS techniques and the tree decoders grows. It is observed that the error probability of the two beam-space tree decoders increases slowly when the number of active users is large. Such an advantage of the beam-space tree decoders mainly comes from enhancing the discriminating power of the decoder by exploiting beam resources.

    \begin{figure}[!htp]
    \begin{center}
    \subfigure[The error probability.]{
    \includegraphics[width=1\linewidth]{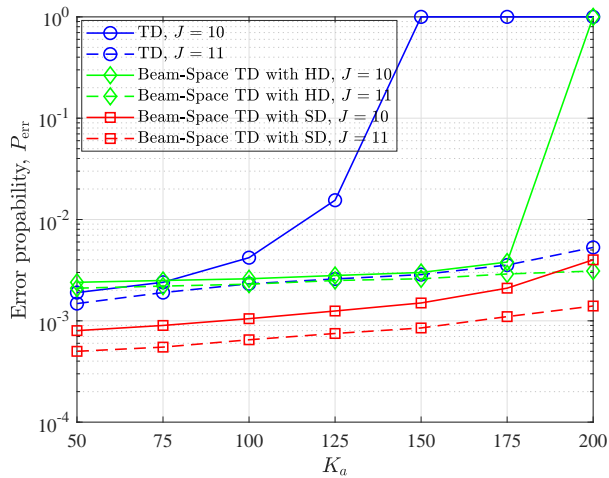}
    \label{ka_perr_11}
    }
    \subfigure[The missed detection rate.]{
    \includegraphics[width=1\linewidth]{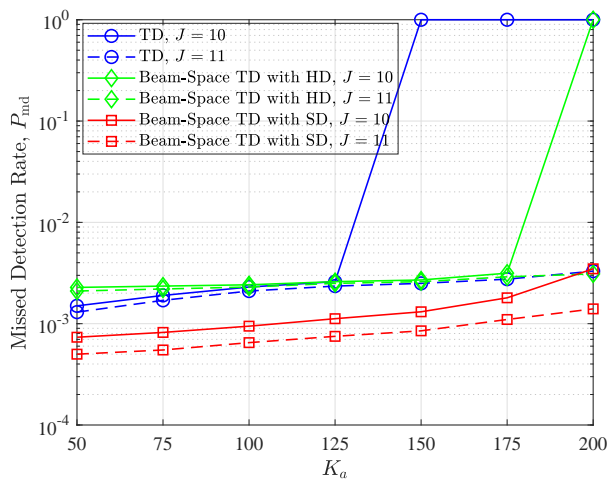}
    \label{ka_perr_11_md}
    }
    \subfigure[The false alarm rate.]{
    \includegraphics[width=1\linewidth]{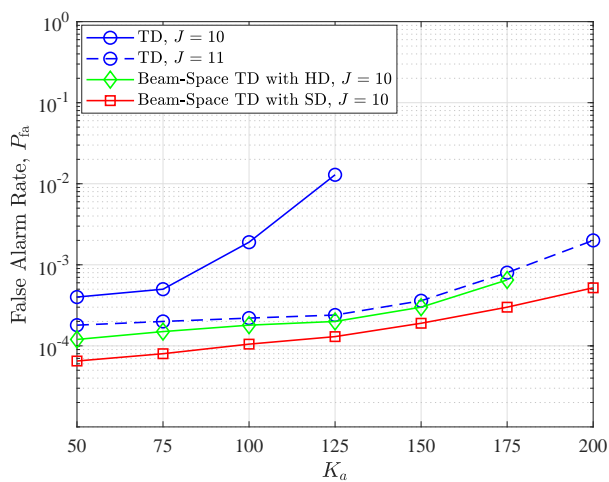}
    \label{ka_perr_11_fa}
    }
    \caption{\rm{The} performance of different URA schemes with different sub-block lengths versus $K_{\mathrm{a}}$ when $E_{\mathrm{b}}/N_0=25\;\rm{dB}$.}
    \label{ka_perr_11_all}
    \end{center}
    \end{figure}

Then we study the impact of the number of parity bits on the system performance. For $J=11$, data profile and parity profile are given as $[11,3,\ldots,3,0,0,0]$ and $[0,8,\ldots$ $,8,11,11,11]$, respectively. In Fig. \ref{ka_perr_11_all}, we compare the performance of the tree decoder (beam domain), the beam-space tree decoder with hard decision, and the beam-space tree decoder with soft decision. It is observed that allocating more parity bits can improve the decoding performance, which is seen in Fig. \ref{ka_perr_11}. Notice that as the number of active users grows, the error probability of the tree decoder and the beam-space tree decoder with hard decision increase sharply to 1. This is because the number of parity bits added to sub-blocks is fixed, the max number of active users the system can serve is limited. Once beyond the limit, the decoders cannot recover any original message of the active users, thus $p_{\mathrm{md}}=1$, $p_{\mathrm{fa}}=0$. As the beam-space tree decoder with hard decision exploits the beam division property, more active users can be served by the system. For the beam-space tree decoder with soft decision, the number of candidate sub-blocks to be kept is determined by the list numbers, not the parity bits. Therefore, the decoder is implemented successfully as the number of active users is sufficiently large. Besides, Fig. \ref{ka_perr_11_md} shows that the missed detection rate of the tree decoder (beam domain) and the beam-space tree decoder with hard decision decreases slowly as the number of parity bits increases. This is because these two algorithms do not deal with the problem of packet loss. Besides, as the number of parity bits is increased, the discriminating power of the tree decoder is improved. Therefore, the false alarm rate of the tree decoder (beam domain) decreases, which is shown in Fig. \ref{ka_perr_11_fa}. Since the beam-space tree decoders exploit beam resources to improve the discriminating power, there is no false alarm when $J=11$. 

\begin{figure}[!htp]
    \small
    \centering
    \includegraphics[width=1\linewidth]{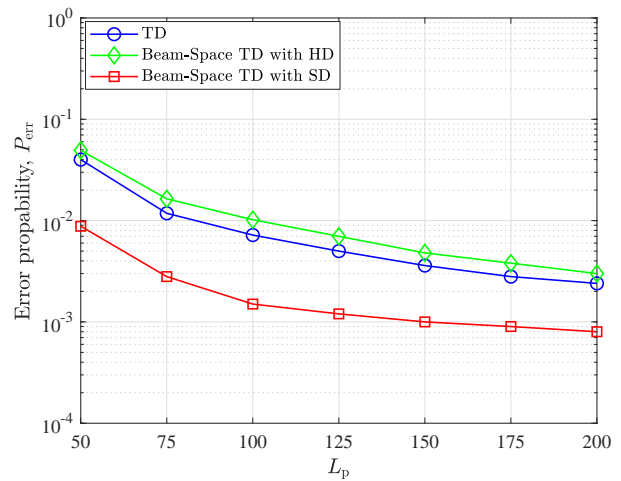}
    \caption{\rm{The} error probability of different URA schemes versus $L_{\mathrm{p}}$.}
    \label{spectral_efficiency}
  \end{figure}
  \begin{figure}[!htp]
    \small
    \centering
    \includegraphics[width=1\linewidth]{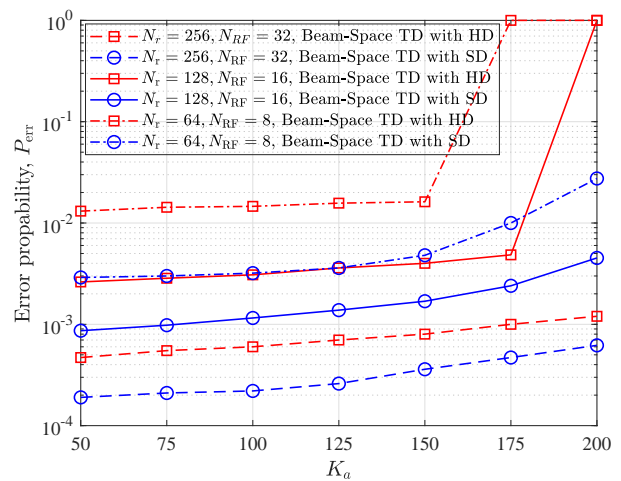}
    \caption{\rm{The} error probability of different URA schemes versus $K_{\mathrm{a}}$ when the BS is equipped with different antennas and RF chains and $E_{\mathrm{b}}/N_0=25\;\rm{dB}$.}
    \label{Nr_Nrf}
    \end{figure}

Fig. \ref{spectral_efficiency} depicts the error probability of different URA schemes with different spectral efficiency when $K_{\mathrm{a}}=50$. The total spectral efficiency is calculated by $\frac{K_{\mathrm{a}}B}{L_{\mathrm{p}}S}$. As the number of observations controls the spectral efficiency, the code length $L_{\mathrm{p}}$ is set as the axis. It can be seen that as the spectral efficiency increases, $L_{\mathrm{p}}$ decreases, and the error probability increases. The reason is that as fewer observations are obtained, the performance of activity detection and channel estimation decreases.

Next, we evaluate the decoding performance with different active users when the BS is equipped with different antennas and RF chains in Fig. \ref{Nr_Nrf}. It can be observed that as the number of antennas and RF chains grows, the error probability decreases, and more active users can be served. The reason is that the BS can generate more narrow beams for the receive beamforming simultaneously, which means that the spatial resolution of the beams is improved, and the beam division property can be exploited completely.

\begin{figure}[!htp]
  \small
  \centering
  \includegraphics[width=1\linewidth]{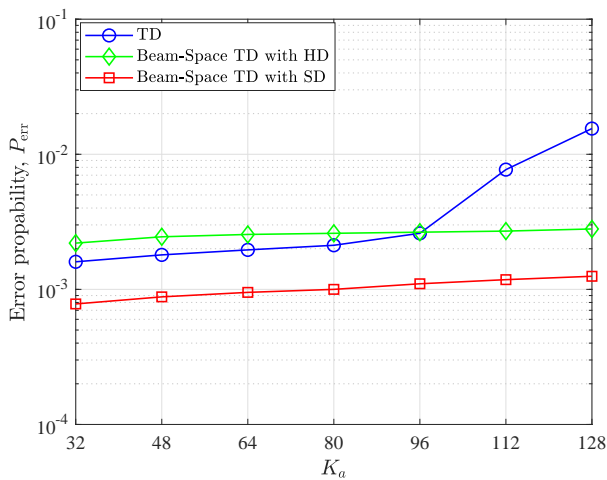}
  \caption{\rm{The} error probability of different URA schemes versus $K_{\mathrm{a}}$ while keeping a fixed ratio between antennas and users.}
  \label{Ka_Nr}
  \end{figure}

Besides, Fig. \ref{Ka_Nr} plots the increasing number of users and antennas together while keep a fixed ratio, i.e., $\frac{K_{\mathrm{a}}}{N_{\mathrm{r}}}=\frac{1}{2}$. It can be observed that as the number of users and antennas increases, the error probability increases slowly. The reason is that although the number of users increases, more antenna array gain is obtained, and the beam division property is exploited. For the conventional tree decoder, as the beam division property is not exploited, the error probability increases more rapidly.

\section{Conclusion}
An URA scheme with beam-space tree decoding under the framework of bit partition and slotted transmission was proposed in this paper. Specifically, we designed two beam-space tree decoders, which are based on hard decision and soft decision, respectively. Both of them exploit the intrinsic beam division property to improve the system performance and help the system serve more active users. Besides, the first decoder can reduce the solution searching space and has low complexity, while the second decoder exploits the advantage of list decoding to recover the miss-detected packets. Simulation results validated that our proposed URA scheme was superior with respect to error probability. 
  
The beam division property is an intrinsic property in mmWave communication systems due to channel propagation. Therefore, this property exploited by our proposed decoders can also be exploited by other schemes under the scenarios of NGMA to help the system serve more users and improve the system performance.
\bibliographystyle{IEEEbib}
\bibliography{ref}

\begin{thebibliography}{10}

\bibitem{bockelmann2016massive}
C.~Bockelmann, N.~Pratas, H.~Nikopour, K.~Au, T.~Svensson, C.~Stefanovic,
  P.~Popovski, and A.~Dekorsy,
\newblock ``Massive machine-type communications in {5G}: Physical and
  {MAC}-layer solutions,''
\newblock {\em IEEE Commun. Mag.}, vol. 54, no. 9, pp. 59--65, 2016.

\bibitem{shariatmadari2015machine}
H.~Shariatmadari, R.~Ratasuk, S.~Iraji, A.~Laya, T.~Taleb, R.~J{\"a}ntti, and
  A.~Ghosh,
\newblock ``Machine-type communications: current status and future perspectives
  toward {5G} systems,''
\newblock {\em IEEE Commun. Mag.}, vol. 53, no. 9, pp. 10--17, 2015.

\bibitem{chen2018sparse}
Z.~Chen, F.~Sohrabi, and W.~Yu,
\newblock ``Sparse activity detection for massive connectivity,''
\newblock {\em IEEE Trans. Signal Process.}, vol. 66, no. 7, pp. 1890--1904,
  2018.

\bibitem{durisi2016toward}
G.~Durisi, T.~Koch, and P.~Popovski,
\newblock ``Toward massive, ultrareliable, and low-latency wireless
  communication with short packets,''
\newblock {\em Proc. IEEE}, vol. 104, no. 9, pp. 1711--1726, 2016.

\bibitem{chen2020massive}
X.~Chen, D.~W.~K. Ng, W.~Yu, E.~G. Larsson, N.~Al-Dhahir, and R.~Schober,
\newblock ``Massive access for {5G} and beyond,''
\newblock {\em IEEE J. Sel. Areas Commun.}, vol. 39, no. 3, pp. 615--637, 2020.

\bibitem{liang2017non}
Y.~Liang, X.~Li, J.~Zhang, and Z.~Ding,
\newblock ``Non-orthogonal random access for {5G} networks,''
\newblock {\em IEEE Trans. Wireless Commun.}, vol. 16, no. 7, pp. 4817--4831,
  2017.

\bibitem{zhang2016grant}
Z.~Zhang, X.~Wang, Y.~Zhang, and Y.~Chen,
\newblock ``Grant-free rateless multiple access: A novel massive access scheme
  for {Internet} of {Things},''
\newblock {\em IEEE Commun. Lett.}, vol. 20, no. 10, pp. 2019--2022, 2016.

\bibitem{senel2018grant}
K.~Senel and E.~G. Larsson,
\newblock ``Grant-free massive {MTC}-enabled massive {MIMO}: A compressive
  sensing approach,''
\newblock {\em IEEE Trans. Commun}, vol. 66, no. 12, pp. 6164--6175, 2018.

\bibitem{kim2019novel}
S.~Kim, H.~Kim, H.~Noh, Y.~Kim, and D.~Hong,
\newblock ``Novel transceiver architecture for an asynchronous grant-free
  {IDMA} system,''
\newblock {\em IEEE Trans. Wireless Commun.}, vol. 18, no. 9, pp. 4491--4504,
  2019.

\bibitem{wang2019joint}
J.~Wang, Z.~Zhang, and L.~Hanzo,
\newblock ``Joint active user detection and channel estimation in massive
  access systems exploiting {Reed--Muller} sequences,''
\newblock {\em IEEE J. Sel. Topics Signal Process.}, vol. 13, no. 3, pp.
  739--752, 2019.

\bibitem{polyanskiy2017perspective}
Y.~Polyanskiy,
\newblock ``A perspective on massive random-access,''
\newblock in {\em IEEE Int. Symp. Inf. Theory (ISIT)}. IEEE, 2017, pp.
  2523--2527.

\bibitem{akdeniz2014millimeter}
M.~R. Akdeniz, Y.~Liu, M.~K. Samimi, S.~Sun, S.~Rangan, T.~S. Rappaport, and
  E.~Erkip,
\newblock ``Millimeter wave channel modeling and cellular capacity
  evaluation,''
\newblock {\em IEEE J. Sel. Areas Commun.}, vol. 32, no. 6, pp. 1164--1179,
  2014.

\bibitem{wen2014channel}
C.-K. Wen, S.~Jin, K.-K. Wong, J.-C. Chen, and P.~Ting,
\newblock ``Channel estimation for massive {MIMO} using {Gaussian}-mixture
  bayesian learning,''
\newblock {\em IEEE Trans. Wireless Commun.}, vol. 14, no. 3, pp. 1356--1368,
  2014.

\bibitem{bellili2019generalized}
F.~Bellili, F.~Sohrabi, and W.~Yu,
\newblock ``Generalized approximate message passing for massive {MIMO} mmwave
  channel estimation with laplacian prior,''
\newblock {\em IEEE Trans. Commun.}, vol. 67, no. 5, pp. 3205--3219, 2019.

\bibitem{sun2015beam}
C.~Sun, X.~Gao, S.~Jin, M.~Matthaiou, Z.~Ding, and C.~Xiao,
\newblock ``Beam division multiple access transmission for massive {MIMO}
  communications,''
\newblock {\em IEEE Trans. Commun.}, vol. 63, no. 6, pp. 2170--2184, 2015.

\bibitem{you2017bdma}
L.~You, X.~Gao, G.~Y. Li, X.-G. Xia, and N.~Ma,
\newblock ``{BDMA} for millimeter-wave/terahertz massive {MIMO} transmission
  with per-beam synchronization,''
\newblock {\em IEEE J. Sel. Areas Commun.}, vol. 35, no. 7, pp. 1550--1563,
  2017.

\bibitem{jia2019massive}
R.~Jia, X.~Chen, Q.~Qi, and H.~Lin,
\newblock ``Massive beam-division multiple access for {B5G} cellular {Internet}
  of {Things},''
\newblock {\em IEEE Internet Things J.}, vol. 7, no. 3, pp. 2386--2396, 2019.

\bibitem{liu2018massive}
L.~Liu and W.~Yu,
\newblock ``Massive connectivity with massive {MIMO—Part I}: Device activity
  detection and channel estimation,''
\newblock {\em IEEE Trans. Signal Process.}, vol. 66, no. 11, pp. 2933--2946,
  2018.

\bibitem{liu2018massive2}
L.~Liu and W.~Yu,
\newblock ``Massive connectivity with massive {MIMO—Part II}: Achievable rate
  characterization,''
\newblock {\em IEEE Trans. Signal Process.}, vol. 66, no. 11, pp. 2947--2959,
  2018.

\bibitem{amalladinne2020coded}
V.~K. Amalladinne, J-F. Chamberland, and K.~R. Narayanan,
\newblock ``A coded compressed sensing scheme for unsourced multiple access,''
\newblock {\em IEEE Trans. Inf. Theory}, vol. 66, no. 10, pp. 6509--6533, 2020.

\bibitem{fengler2021sparcs}
A.~Fengler, P.~Jung, and G.~Caire,
\newblock ``{SPARCs} for unsourced random access,''
\newblock {\em IEEE Trans. on Inf. Theory}, pp. 1--1, 2021.

\bibitem{fengler2021non}
A.~Fengler, S.~Haghighatshoar, P.~Jung, and G.~Caire,
\newblock ``Non-{Bayesian} activity detection, large-scale fading coefficient
  estimation, and unsourced random access with a massive {MIMO} receiver,''
\newblock {\em IEEE Trans. Inf. Theory}, vol. 67, no. 5, pp. 2925--2951, 2021.

\bibitem{shyianov2020massive}
V.~Shyianov, F.~Bellili, A.~Mezghani, and E.~Hossain,
\newblock ``Massive unsourced random access based on uncoupled compressive
  sensing: Another blessing of massive {MIMO},''
\newblock {\em IEEE J. Sel. Areas Commun.}, vol. 39, no. 3, pp. 820--834, 2020.

\bibitem{decurninge2020tensor}
A.~Decurninge, I.~Land, and M.~Guillaud,
\newblock ``Tensor-based modulation for unsourced massive random access,''
\newblock {\em IEEE Wireless Commun. Lett.}, vol. 10, no. 3, pp. 552--556,
  2020.

\bibitem{fengler2020pilot}
A.~Fengler, P.~Jung, and G.~Caire,
\newblock ``Pilot-based unsourced random access with a massive {MIMO} receiver,
  {MRC} and polar codes,''
\newblock {\em arXiv preprint arXiv:2012.03277}, 2020.

\bibitem{CAD}
X.~Shao, X.~Chen, D.~W.~K. Ng, C.~Zhong, and Z.~Zhang,
\newblock ``Cooperative activity detection: Sourced and unsourced massive
  random access paradigms,''
\newblock {\em IEEE Trans. Signal Process.}, vol. 68, pp. 6578--6593, 2020.

\bibitem{gao2016near}
X.~Gao, L.~Dai, Z.~Chen, Z.~Wang, and Z.~Zhang,
\newblock ``Near-optimal beam selection for beamspace mmwave massive {MIMO}
  systems,''
\newblock {\em IEEE Commun. Lett.}, vol. 20, no. 5, pp. 1054--1057, 2016.

\bibitem{wang2017spectrum}
B.~Wang, L.~Dai, Z.~Wang, N.~Ge, and S.~Zhou,
\newblock ``Spectrum and energy-efficient beamspace {MIMO-NOMA} for
  millimeter-wave communications using lens antenna array,''
\newblock {\em IEEE J. Sel. Areas Commun.}, vol. 35, no. 10, pp. 2370--2382,
  2017.

\bibitem{donoho2009message}
D.L. Donoho, A.~Maleki, and A.~Montanari,
\newblock ``Message-passing algorithms for compressed sensing,''
\newblock {\em Proc. Nat. Acad. Sci.}, vol. 106, no. 45, pp. 18914--18919,
  2009.

\bibitem{vila2013expectation}
J.P. Vila and P.~Schniter,
\newblock ``Expectation-maximization {Gaussian}-mixture approximate message
  passing,''
\newblock {\em IEEE Trans. Signal Process.}, vol. 61, no. 19, pp. 4658--4672,
  2013.

\bibitem{idma}
P.~Li, L.~Liu, K.~Wu, and W.K. Leung,
\newblock ``Interleave division multiple-access,''
\newblock {\em IEEE Trans. Wireless Commun.}, vol. 5, no. 4, pp. 938--947,
  2006.

\bibitem{tal2015list}
I.~Tal and A.~Vardy,
\newblock ``List decoding of polar codes,''
\newblock {\em IEEE Trans. Inf. Theory}, vol. 61, no. 5, pp. 2213--2226, 2015.

\bibitem{fengler2019massive}
A.~Fengler, G.~Caire, P.~Jung, and S.~Haghighatshoar,
\newblock ``Massive {MIMO} unsourced random access,''
\newblock {\em arXiv preprint arXiv:1901.00828}, 2019.

\end{thebibliography}
\end{document}